\documentclass[oldversion]{aa}
\usepackage{txfonts}
\usepackage{natbib}
\usepackage{subfigure}
\bibpunct{(}{)}{;}{a}{}{,} 
\usepackage{graphicx}
\usepackage{amssymb}
\usepackage{threeparttable}
\usepackage{rotating}
\usepackage{stfloats}
\usepackage{caption}
\usepackage{booktabs}

\usepackage{epstopdf}

\newcommand{\Msun}{\mathrm{M}_{\odot}}

\newcommand{\lum}{\mathrm{erg~s}^{-1}}
\newcommand{\flux}{\mathrm{erg~cm}^{-2}~\mathrm{s}^{-1}}
\newcommand{\cnts}{\mathrm{c~s}^{-1}}
\newcommand{\nh}{\mathrm{cm}^{-2}}

\newcommand{\asca}{\textit{ASCA}}
\newcommand{\einstein}{\textit{Einstein}}
\newcommand{\swift}{\textit{Swift}}
\newcommand{\chan}{\textit{Chandra}}
\newcommand{\xmm}{\textit{XMM-Newton}}
\newcommand{\inte}{\textit{INTEGRAL}}
\newcommand{\rosat}{\textit{ROSAT}}

\newcommand{\ucxb}{AX J1754.2--2754}
\newcommand{\porb}{AX J1538.3--5541}
\newcommand{\xte}{XTE J1719--291}
\newcommand{\polar}{AX J1620.1--5002}

\begin{document}

\title{Swift follow-up observations of unclassified ASCA sources}

\titlerunning{Swift observations of unclassified ASCA sources}

\author{
N. Degenaar\inst{1,2}\thanks{e-mail: degenaar@umich.edu} \fnmsep \thanks{Hubble fellow}
\and R.L.C. Starling\inst{3}
\and P.A. Evans\inst{3}
\and A.P. Beardmore\inst{3}
\and D.N. Burrows\inst{4}
\and E. M. Cackett\inst{5}
\and S. Campana\inst{6}
\and \\ D. Grupe\inst{4}
\and J. Kennea\inst{4}
\and A. Moretti\inst{6}
\and C. Pagani\inst{3}
\and K.L. Page\inst{3}
\and V. La Parola\inst{7}
\and R.Wijnands\inst{2}
}

\institute{
University of Michigan, Department of Astronomy, 500 Church St, Ann Arbor, MI 48109, USA
\and Astronomical Institute ``Anton Pannekoek", University of Amsterdam, Postbus 94249, 1090 GE, Amsterdam, The Netherlands
\and Department of Physics and Astronomy, University of Leicester, University Road, Leicester LE1 7RH, UK
\and Department of Astronomy and Astrophysics, 525 Davey Lab, Pennsylvania State University, University Park, PA 16802, USA 
\and Institute of Astronomy, University of Cambridge, Madingley Road, Cambridge CB3 0HA, UK
\and INAF-Osservatorio Astronomico di Brera, Via Bianchi 46, I-23807 Merate (Lc), Italy
\and INAF-IASF Palermo, Via Ugo La Malfa 153, 90146 Palermo, Italy
}

\authorrunning{N. Degenaar et al.}

\date{Received 13 December 2011 / Accepted 29 January 2012}

\abstract {
We present \swift\ follow-up observations of a sample of 35 unclassified faint X-ray sources drawn from the \asca\ Galactic centre and Galactic plane surveys. Our short, pointed XRT observations allow detections down to a limiting 0.3--10 keV flux of $F_X \sim10^{-13}~\flux$, which translates into a luminosity of $L_X\sim10^{33}~\lum$ for an assumed distance of $D=8$~kpc. The brightest source in our sample reaches a maximum 0.3--10 keV luminosity of $L_X\sim2\times10^{36}~(D/8~\mathrm{kpc})^2~\lum$ during our \swift\ observations. We detect 16 ($46\%$) of the \asca\ sources with the XRT, while 19 were not detected during our program. Since we are probing the faint end of the \asca\ source populations, we expect a large fraction of the non-detections to be due to the Eddington bias. This is strengthened by the fact that we find the observed XRT count rates to be predominantly lower than expected based on the reported \asca\ intensities. Nevertheless, investigation of the \asca\ properties and any possible long-term X-ray variability leads us to conclude that the non-detections likely include two spurious \asca\ detections and three objects that are highly variable or transient X-ray sources. For the 16 XRT-detected sources we obtain positional accuracies of $\sim2-4''$, which significantly improves upon their \asca\ uncertainties of $\sim1'$. We use the X-ray spectra and variability to characterise these objects. Most appear to be faint, persistent X-ray emitters that have highly absorbed spectra. Based on their X-ray properties we identify three accreting compact objects: one confirmed (\ucxb) and one candidate (\porb) X-ray binary, and one possible magnetically accreting white dwarf (\polar). Furthermore, we use the improved positions of XRT-detected sources to search for counterparts in simultaneously obtained \swift/UVOT images and possible associations with catalogued sources at various wavelengths. This reveals three possible main sequence stars amongst our sample. The other sources remain unclassified, but our improved XRT positions provide good prospects for dedicated follow-up observations that have the potential to elucidate the nature of these faint, unclassified \asca\ sources.
}


\maketitle

\section{Introduction}
The Advanced Satellite for Cosmology and Astrophysics (\asca) was launched in 1993 and operated until 2000. It was the first mission that combined imaging capabilities with a relatively broad energy passband (0.7--10 keV). During its lifetime, the \asca\ satellite pointed towards the Galactic plane and the inner bulge of the Galaxy many times. In particular, two dedicated surveys were performed that detected about 200 X-ray sources with 0.7--10 keV fluxes of $F_X\gtrsim 3\times 10^{-13}~\flux$ \citep[][]{sugizaki01,sakano02}. Both surveys used the GIS (gas imaging spectrometer) detector, which had a spatial resolution of $\sim30\arcsec$ and a field of view (FOV) of $\sim50'$.

The \asca\ Galactic plane survey entailed a systematic study of X-ray sources located in the inner Galactic disk \citep{sugizaki01}. The region of $| l | \lesssim45^{\circ}$, $| b |\lesssim 0.4^{\circ}$ was covered between 1996 and 1999 using successive 10-ks observations. Investigation of three different energy bands (0.7--2, 2--10 and 0.7--7 keV) resulted in the identification of 163 individual X-ray sources detected with a significance of $\gtrsim 4 \sigma$. \asca\ targeted a region of $5^{\circ} \times 5^{\circ}$ around the centre of our Galaxy between 1993 and 1999 \citep{sakano02}. Searching the 0.7--3 and 3--10 keV images yielded a total of 52 point sources that were detected with a significance of $\gtrsim5\sigma$. In both \asca\ surveys, spectral analysis was performed for sources that were detected with a significance of $\gtrsim5\sigma$. The GIS data were fitted to a simple absorbed powerlaw model to characterise the X-ray spectra \citep[][]{sugizaki01,sakano02}. Source positions could be determined with a $90\%$ confidence positional accuracy of $\sim1'$ \citep{ueda1999,sugizaki01,sakano02}.

After cross-correlation with the \einstein\ and \rosat\ bright source catalogues, about 2/3 of the \asca-detected X-ray sources remained unclassified \citep{sugizaki01,sakano02}. These objects have 0.7--10 keV X-ray fluxes in the range of $F_X \sim 10^{-13} - 10^{-11}~\flux$, which translates into a luminosity of $L_X \sim 10^{33}-10^{35}~\lum$ for a distance of $D=8$~kpc. This is right in between the well-known bright X-ray point sources ($L_X>10^{36}~\lum$) that are mostly active X-ray binaries \citep[e.g.,][]{chen97}, and dim X-ray sources ($L_X<10^{33}~\lum$) that consist of a collection of source types such as accreting white dwarfs, dormant X-ray binaries, active stars or young pulsars \citep[e.g.,][]{muno2009}. 

The luminosity range traced by the unclassified \asca\ sources suggests that a variety of source types might be amongst them. For example, it covers the intensities typically seen for strongly magnetised neutron stars (magnetars), bright magnetically accreting white dwarfs (polars and intermediate polars), sub-luminous X-ray binaries and X-ray emitting massive stars \citep[e.g.,][]{maeda1996,ezuka1999,wijnands06,chelovekov07,gelfland2007,heras08,kaur2010,degenaar09_gc,anderson2011}. In addition, the sample may contain foreground stars or background active galactic nuclei (AGN). 
The main difficulty in classifying the \asca\ sources is their relatively large positional uncertainties of $\sim$1$'$. This inhibits follow-up observations aiming to search for counterparts at other wavelengths.

We launched a program to observe unclassified \asca\ sources with the \swift\ satellite \citep[][]{gehrels2004}. The primary instrument for our observations is the X-ray telescope \citep[XRT;][]{burrows05}. The XRT is sensitive in the 0.3--10 keV energy range and has a FOV of $\sim23' \times 23'$. Furthermore, we used
data obtained with the Ultra-Violet/Optical Telescope \citep[UVOT;][]{roming05} that has a FOV of $\sim17' \times 17'$ and can be operated using a variety of filters in a wavelength range of $\sim1500-8000$ \AA{} \citep[][]{poole2008}. In addition to these narrow-field instruments, \swift\ is equipped with the Burst Alert Telescope \citep[BAT;][]{barthelmy05}: a hard X-ray (15--150 keV) monitoring instrument that has a wide FOV ($2$~steradian). The combination of its pointing flexibility, good sensitivity in the soft X-ray band (below 10 keV) and multi-wavelength coverage makes \swift\ an ideal tool to perform follow-up observations of X-ray sources discovered by other missions \citep[e.g.,][]{rodriguez09,starling2011,kennea2011}.

By observing unclassified \asca\ sources with \swift, our primary aim is to improve their positional uncertainties to a few arcseconds, so that dedicated searches for counterparts at other wavelengths become feasible. In addition, we strive to obtain X-ray spectral information and to study possible variability in their X-ray emission. This approach provides a basis for further investigation of the nature of individual sources, as well as the population of faint \asca\ sources as a whole.\footnote{See also the {\it ChIcAGO} project, which targets a (different) sample of unclassified \asca\ sources using \chan\ and multi-wavelength follow-up observations \citep[][]{anderson2011}.}

\section{Sample selection and observations}\label{sample}
In 2006, we compiled a target list of 48 unclassified \asca\ sources drawn from the surveys of \citet[][]{sugizaki01} and \citet{sakano02}. We selected these targets because an accurate position was not reported at the time and they were not previously (or scheduled to be) observed with \chan\ or \xmm, which would also provide an accurate localisation \citep[see][]{cackett06,anderson2011}.

Between 2006 November 3 and 2010 November 4, \swift\ observed 35 of our targets through the fill-in program (i.e., $73\%$ of our initial sample). Table~\ref{tab:obs} presents the observation log of these sources. Typically, the total exposure time per source was about $5$~ks, although considerably longer exposures were obtained on some occasions (see Table~\ref{tab:obs}). The XRT was operated in the Photon Counting (PC) mode \citep[][]{hill2004} and the UVOT in the `filter of the day' mode, so a variety of optical/UV filters was used in this program. 

\begin{table}
\begin{threeparttable}[h!]
\begin{center}
\caption{Target list and \swift/XRT observation log.}
\begin{tabular}{l c c c c c} 
\toprule
$\#$ & Name & \swift\ ID & Obs. date & $t_{\mathrm{exp}}$ &  Cat. \\
 &  &  & & (ks) &  \\
\midrule
1 & AX J1457.5--5901 &  36134  & & 4.8 &  1\\
& &  & 2007-01-24  & 3.3 &  \\
& &  & 2007-01-28  & 1.5 &   \\ 
2* & AX J1504.6--5824 & 36135 & 2007-01-24 & 5.3 &  1 \\
3 & AX J1510.0--5824 & 36136 & 2007-01-24 & 4.7 &  1 \\
4 & AX J1537.8--5556 & 36137 & & 5.6 &  1 \\
 &  & & 2007-01-07 & 4.3 &   \\
 & &  & 2007-01-11 & 1.3 &   \\
5* & \porb\ & 36138 & & 89.9 & 1 \\
& & & 2007-01-05 & 2.2 &   \\
& & & 2007-01-31 & 0.9 &    \\
& & & 2007-03-16 & 2.8 &     \\
& & & 2007-03-17 & 2.7 &     \\
& & & 2007-08-22 & 0.6 &    \\
& & & 2007-08-30 & 2.3  &     \\
& & & 2007-08-31 & 15.1 &    \\
& & & 2007-09-01 & 15.5 &    \\
& & & 2007-09-03 & 13.8 &   \\
& & & 2007-09-06 & 0.4 &    \\
& & & 2007-09-07 & 5.3 &    \\
& & & 2007-09-08 & 10.7 &     \\
& & & 2007-09-10 & 5.4 &     \\
& & & 2007-09-11 & 5.3 &     \\
& & & 2007-09-13 & 6.9 &     \\
6 & AX J1545.9--5443 & 36139 & 2007-01-21 & 4.7 &  1\\
7* & \polar\ & 36140 & 2007-02-07 & 4.7 & 1 \\
8* & AX J1651.0--4403 & 36141  & 2008-01-21  & 4.9 & 1\\
9 & AX J1657.3--4321 & 36142 & 2008-01-22 & 4.9 &  1\\
10 & AX J1659.8--4209 &  36143 & 2008-01-23 & 2.0 &  1 \\
11* & AX J1717.2--3718 & 36145  & 2009-02-03 & 1.7 &  1\\
12* & AX J1719.3--3703 & 36146  & & 4.3 & 1\\
& & & 2007-04-07 & 1.4 &   \\
& & & 2008-04-29 & 1.2 &    \\
& & & 2008-07-01 & 1.7 &    \\
13* & AX J1720.8--3710  & 36147  & 2009-01-30 &1.5 &  1\\
14* & AX J1725.8--3533  & 36148  & & 5.6 &  1\\
& & & 2007-06-19 & 1.9 &   \\
& & & 2009-01-31 & 3.7 &   \\
15 & AX J1734.5--2915 & 36149  & 2008-02-16 & 6.1 &  2 \\
16* & AX J1738.4--2902  &  36151 & 2007-10-28 & 4.9 & 2 \\
17* & AX J1739.5--2730  &  36153 & 2008-02-08 & 1.7 &  2\\
18* & AX J1742.6--3022  &  36156 & 2008-03-07 & 4.8 &  2\\
19* & AX J1742.6--2901  &  36157 & 2008-03-07 & 6.5 &  2\\
20 & AX J1751.1--2748  & 36159  & & 4.7 &  2\\
& & & 2007-03-18 & 3.5 &  \\
& & & 2009-02-17 & 1.2 &    \\
21 & AX J1753.5--2745  &  36160 & 2008-07-11 & 4.1 & 2 \\
22* & \ucxb\ & 36163  & & 16.3 &  2 \\
& & & 2007-07-12  & 2.9	 &  \\
& &  & 2007-07-19  & 2.9	 &  \\
& &  & 2008-07-18  & 2.0	 &  \\
& &  & 2008-07-20  & 1.1 & 	 \\
& &  & 2008-07-22  & 0.5 & 	 \\
& &  & 2008-07-24  & 1.2	 &  \\
& &  & 2008-07-26  & 1.9	 &  \\
& &  & 2008-07-31  & 1.0	 &  \\
& &  & 2008-07-31  & 1.0 & 	 \\
& &  & 2008-08-02  & 1.0 &  \\
& &  & 2008-08-02  & 0.8 & 	 \\
\bottomrule
\end{tabular}
\label{tab:obs}
\begin{tablenotes}
\item[]Note. -- Catalogues: 1=\citet{sugizaki01}, 2=\citet{sakano02}. Sources marked by an asterisk were detected with \swift/XRT, while the others were not. If multiple observations were performed, we also list the total exposure time on the source.
\end{tablenotes}
\end{center}
\end{threeparttable}
\end{table}

\begin{table}
\ContinuedFloat
\begin{threeparttable}[h!]
\begin{center}
\caption{Continued}
\begin{tabular}{l c c c c c} 
\toprule
$\#$ & Name & \swift\ ID & Obs. date & $t_{\mathrm{exp}}$ & Cat. \\
 &  &  & & (ks) &  \\
\midrule
23 & AX J1758.0--2818  & 36167  & & 2.5 &  2 \\
& & &  2008-05-07 & 1.0 &     \\
& & & 2010-02-05 & 1.2 &     \\
& & & 2010-11-04 & 0.3 &    \\
24 & AX J1832.5--0916  &  36174 & 2008-02-29 & 2.4 &  1\\
25 & AX J1833.9--0822 & 36175  & 2007-11-15 & 4.4 &   1\\
26 & AX J1834.6--0801 & 36176  & & 6.4 &  1 \\
& & & 2007-03-04 & 3.1 &    \\
 & &  & 2008-02-26 & 3.3 &    \\
27 & AX J1835.1--0806  & 36177  & & 7.1 &  1\\
& & & 2007-03-28 & 5.3 &    \\
&  &  & 2007-04-02 & 1.8 &    \\
28 & AX J1836.3--0647  & 36178  & & 7.7 &   1\\
& & & 2007-03-09 & 3.2 &    \\
& & & 2007-03-20 & 4.5 &   \\
29* & AX J1846.0--0231 & 36179  & & 12.9  &1 \\
& & & 2006-11-03 & 4.6  &   \\
& & (36180)  & 2006-11-10 & 4.7 &   \\
& & (36180)  & 2007-03-02 & 3.6 &    \\
30* & AX J1846.1--0239 & 36180  & & 12.9 & 1\\
& & & 2006-11-10 & 4.7 &    \\
& & & 2007-03-02 & 3.6 &   \\
& & (36179) & 2006-11-03 & 4.6 &  \\
31 & AX J1847.6--0219  & 36181  & & 7.3 &  1 \\
& &  & 2007-03-01 & 1.8 &    \\
&  &  & 2007-03-02  & 5.1 &    \\
&  &  & 2007-03-16  & 0.4 &    \\
32* & AX J1856.7+0220  &  36182 & & 5.4 &  1\\
&&  & 2006-10-25 & 4.5 &  \\
&   &  & 2006-11-10 & 1.0 &   \\
33 & AX J1856.8+0245  &  36183 & & 11.2 & 1 \\
& & & 2006-11-10 & 0.7 &    \\
& & & 2007-03-13 & 4.1 &   \\
& &  (36184) & 2007-03-10 & 4.1 &   \\
34 & AX J1857.3+0247  &  36184 & & 11.2 &  1 \\
& & & 2007-03-07 & 3.1 &   \\
& & & 2007-03-10 & 4.1 &   \\
& & (36183) & 2007-03-13 & 4.1 &   \\
35 & AX J1900.1+0427 &  36185 & 2006-11-04 & 3.8 &  1\\
\bottomrule
\end{tabular}
\begin{tablenotes}
\item[]Note. -- AX J1846.0--0231 and AX J1846.1--0239 lie in each others FOV, as do AX J1856.8+0245 and AX J1857.3+0247. For these sources we combined the data sets in our analysis.
\end{tablenotes}
\end{center}
\end{threeparttable}
\end{table}

\section{Data analysis}\label{obs_ana}

All {\it Swift} data were reduced using the \swift\ tools v. 3.8 within \textsc{heasoft} v. 6.11. The latest calibration data as of 2011 August (\textsc{caldb} v. 3.8) were used. Source detection and position determination were carried out using the recipes described in \citet{evans09}. When possible, X-ray positions were calculated using the UVOT-enhancement method, matching field sources detected by the UVOT to the USNO-B1 catalogue \citep{goad07,evans09}. We initially searched for X-ray sources that were detected in the 0.3--10 keV XRT images with a significance of $3\sigma$ above the background, and were located within $90''$ of the reported \asca\ coordinates \citep[corresponding to the $\sim2.5\sigma$ positional uncertainties;][]{sugizaki01,sakano02}.\footnote{We note that due to erroneous star tracker readings the \asca\ positional uncertainties were typically larger than expected. Corrections to apply to SIS and some of the GIS data have been calculated in \citet{gotthelf2000}, however we choose to adopt the positional accuracies reported in \citet{sugizaki01} and \citet{sakano02} as a conservative approach.} We next visually inspected all XRT images to search for possible additional detections.

We extracted light curves and spectra from the XRT data for all detected sources, using the methods outlined in \citet{evans09}. To provide a characterisation of the X-ray spectrum and to allow for a direct comparison with the \asca\ results, the spectral data were fitted with a simple absorbed powerlaw model (PHABS*POWERLAW) using \textsc{xspec} (v. 12.7). The PHABS model accounts for the neutral hydrogen absorption along the line of sight, for which we used the default \textsc{xspec} abundances \citep[][]{anders1989_phabs_abun} and cross-sections \citep[][]{balucinska1992_phabs_cross}. 

Given that our targets generally had low count rates and the XRT exposure times were typically relatively short, we grouped the spectra to contain at least one photon per bin and employed Cash statistics \citep[][]{cash1979}. We calculated the absorbed and unabsorbed flux for all detected sources in the full XRT energy band (0.3--10 keV). The unabsorbed flux was converted into a 0.3--10 keV luminosity assuming a fiducial distance of $D=8$~kpc, unless stated otherwise. For two sources with long exposures, \ucxb\ and \porb, we obtained enough statistics to allow for a more in-depth investigation of the X-ray spectrum (see Sect.~\ref{subsubsec:ucxb} and~\ref{subsubsec:porb}). 

In the case of non-detections, we used the inferred $3\sigma$ upper limits on the source count rates to estimate the corresponding upper limits on the 0.3--10 keV fluxes and luminosities by employing \textsc{pimms} (v. 4.4). Our non-detections concern mostly sources that were detected only with low significance during the \asca\ surveys ($\lesssim4\sigma$; see Sect.~\ref{subsec:nat_nondetect}) and hence spectral information was not available \citep[][]{sugizaki01,sakano02}. We therefore assumed a fiducial powerlaw spectral model with an index of $\Gamma=1.9$ (the average value from fitting the XRT spectra of detected sources; see Sect.~\ref{subsec:detected}), and used a hydrogen column density set to the Galactic value in the direction of a source as determined in the Leiden-Argentine-Bonn (LAB) Galactic neutral hydrogen survey \citep[$N_{H}^{\mathrm{Gal}}$;][]{kalberla2005}. 

To study possible long-term variability amongst our targets, we employed \textsc{pimms} to estimate the expected XRT count rates based on the intensities measured by \asca\ \citep[][]{sugizaki01,sakano02}. We compared these values with the actual observed count rates (or upper limits) to find any indications of strong X-ray variability between the time of the \asca\ surveys (1993--1999) and our XRT follow-up program (2006--2010). This also aided in evaluation of any possible spurious \asca\ detections (see Sect.~\ref{subsec:nat_nondetect}). Whenever spectral information was available from the \asca\ data, we used the reported hydrogen column density ($N_H$), powerlaw spectral index ($\Gamma$) and 0.7--10 keV unabsorbed flux to estimate the source count rates expected for our \swift/XRT PC-mode observations. If no spectral information was available from the \asca\ surveys, we estimated the expected XRT count rates by converting the reported \asca\ count rates (0.7--7 keV, or 2--10 keV when the former was not available), assuming a fiducial powerlaw spectrum with $\Gamma=1.9$ and using $N_{H}^{\mathrm{Gal}}$.

For the detected sources we searched the simultaneously obtained UVOT images for possible UV/optical counterparts within the $90\%$ confidence XRT error circle. Where a tentative counterpart could be identified, we obtained the UVOT magnitudes using the {\it Swift} tool \textsc{uvotsource}. In addition, we cross-correlated the improved X-ray positions of our targets with a variety of astronomical data bases using {\it Vizier} (e.g., {\it NOMAD}, {\it USNO-B1}, {\it 2MASS}) in order to assess any possible associations of the \asca\ sources with catalogued objects.

\begin{figure*}
 \begin{center}
\includegraphics[width=8.5cm]{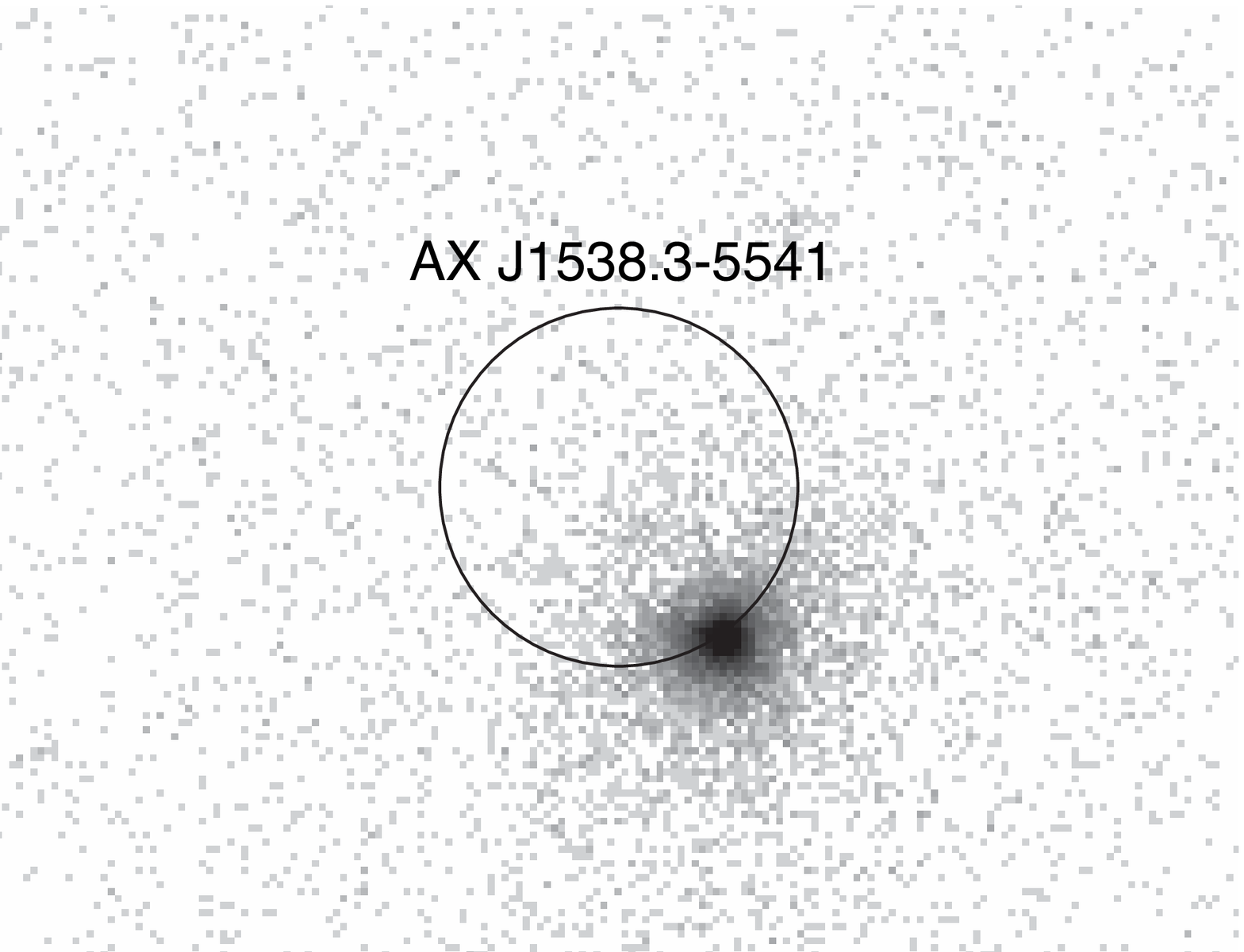}\hspace{0.2cm}
\includegraphics[width=8.5cm]{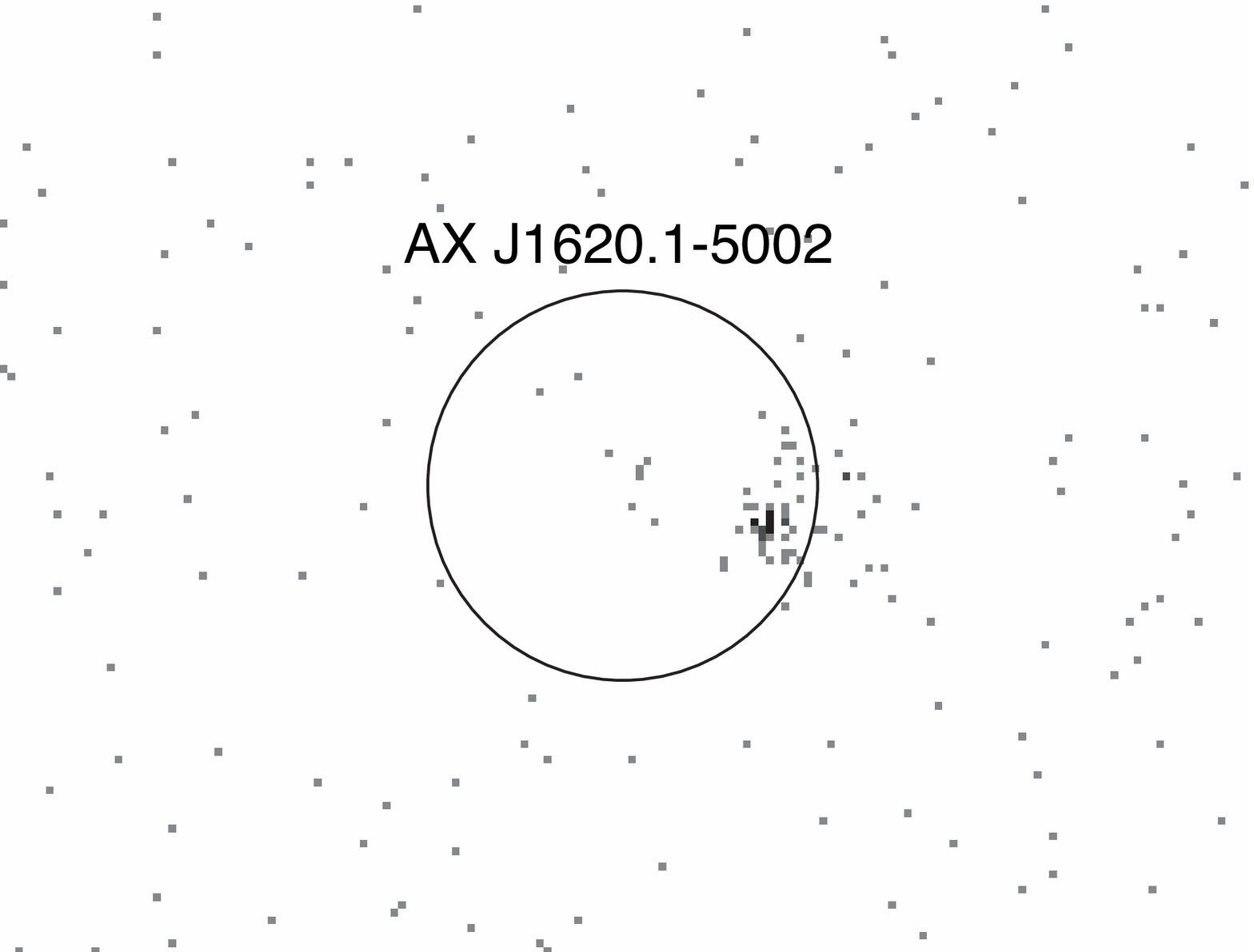}
    \end{center}
\caption[]{\swift/XRT 0.3--10 keV images of two detected sources. The circles represent the ASCA positional uncertainties of $1'$. For illustration we picked one relatively bright (left) and one relatively faint (right) object from our sample of detected sources.
}
 \label{fig:images}
\end{figure*}

\begin{figure}
 \begin{center}
\includegraphics[width=8.5cm]{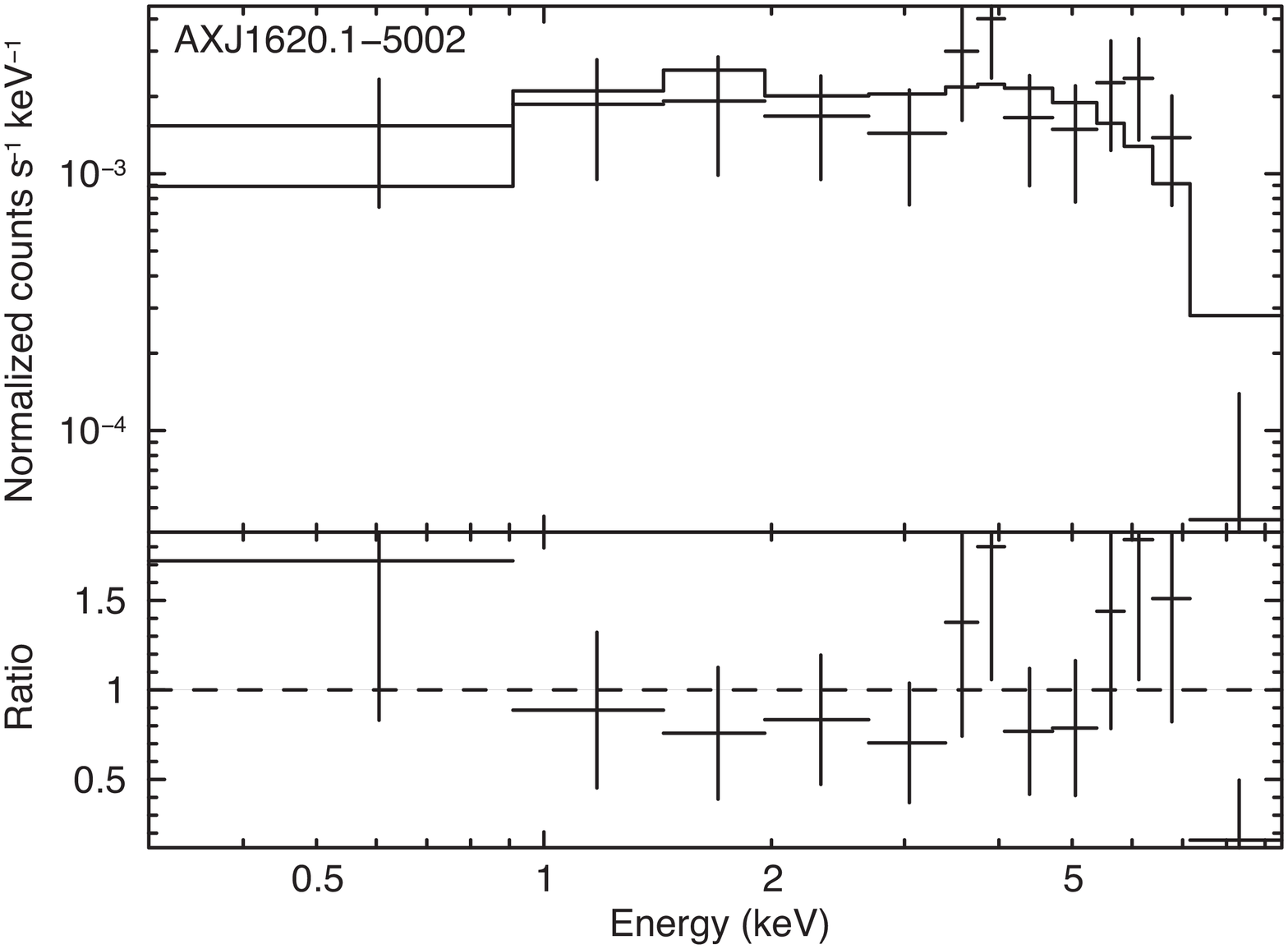}
    \end{center}
\caption[]{\swift/XRT spectrum of \polar\ (rebinned to a signal to noise ratio of 5 for display purposes). The solid line represents the best-fit result to an absorbed powerlaw model. The bottom panel plots the data to model ratios. 
}
 \label{fig:wd}
\end{figure}

\section{Results}\label{results}

\subsection{X-ray detected sources}\label{subsec:detected}
Out of the 35 targets observed with \swift/XRT, we detect 11 X-ray sources at $\gtrsim3\sigma$ significance within our $90''$ search region centred on the \asca\ coordinates. We thus have firm detections for $31\%$ of our observed targets. Their inferred XRT positions are listed in Table~\ref{tab:pos}. The mean XRT positional uncertainty for these objects is $3.2''$, which is a considerable improvement over their \asca\ uncertainties of $1'$ (both quoted errors refer to 90\% confidence levels). In Fig.~\ref{fig:images} we display two example XRT images of detected sources, indicating the \asca\ positional uncertainties. Querying catalogues of known X-ray sources suggests that the probability of detecting an unrelated X-ray source within our $90''$ search region is relatively low (between $0.5-5\%$ for our detected sources).

In Table~\ref{tab:spec}, we present the results of our spectral analysis of the XRT-detected sources. All uncertainties on spectral parameters quoted in this work refer to 90\% confidence levels. Fitting the spectral data with a simple absorbed powerlaw model generally provided a reasonable description of the data. Most detected sources have highly absorbed X-ray spectra ($N_H\gtrsim 1\times10^{22}~\nh$). The hydrogen column densities inferred from our spectral analysis are typically higher than the estimated Galactic values ($N_H^{\mathrm{Gal}}$) in the direction of the sources (see Table~\ref{tab:spec}). We find an average column density of $N_H= 3.6\times10^{22}~\nh$ and photon indices that are typically in the range of $\Gamma \sim1.5-3$. The average powerlaw spectral index inferred for our detected sources is $\Gamma=1.9$.

\setcounter{table}{1}
\begin{table*}
\begin{threeparttable}[t]
\begin{center}
\caption[]{{Positions (J2000) and errors for detected sources.}}
\begin{tabular}{l l c c c c}
\toprule
$\#$ & Name & R.A. & Dec. & Error & Offset \\
& & (h m s) & $(^{\circ}~'~'')$ & (\arcsec) & (\arcsec) \\
\midrule
\multicolumn{6}{l}{{\bf Firm \swift/XRT detections}} \\ 
2 & AX J1504.6--5824 & 15 04 30.92 & --58 24 09.6 & 4.1 & 55.3 \\
5 & AX J1538.3--5541 & 15 38 13.68 & --55 42 13.1 & 1.9 & 62.2 \\
7 & \polar & 16 20 06.76 & --50 01 58.0 & 2.8 & 47.2 \\
8 & AX J1651.0--4403 & 16 51 11.43 & --44 02 37.0 & 2.3 & 79.8 \\
12 & AX J1719.3--3703 & 17 19 20.88 & --37 01 51.6 & 4.4 & 72.9 \\
14 & AX J1725.8--3533 & 17 25 49.31 & --35 33 40.1 & 4.1 & 15.7 \\
16 & AX J1738.4--2902$\dagger$ & 17 38 26.18 & --29 01 49.6 & 2.2 & 3.0 \\
18 & AX J1742.6--3022 & 17 42 41.13 & --30 22 36.8 & 2.8 & 75.9 \\
19 & AX J1742.6--2901 & 17 42 41.93 & --29 02 09.7 & 3.4 & 5.6 \\
22 & AX J1754.2--2754$\dagger$ & 17 54 14.47 & --27 54 36.1 & 2.2 & 48.7 \\
29 & AX J1846.0--0231 & 18 45 56.81 & --02 31 59.3 & 4.4 & 60.9 \\
\midrule
\multicolumn{6}{l}{{\bf Tentative \swift/XRT detections}} \\ 
11 & AX J1717.2--3718 & 17 17 24.01 & --37 17 18.3 & 3.2 & 105 \\
13 & AX J1720.8--3710 & 17 20 51.64 & --37 10 33.9 & 3.9 & 23.7 \\
17 & AX J1739.5--2730 & 17 39 35.86 & --27 29 36.8 & 4.2 & 40.2 \\
30 & AX J1846.1--0239 & 18 46 03.75 & --02 39 16.7 & 4.1 & 106 \\
32 & AX J1856.7+0220 & 18 56 43.82 & +02 19 20.15 & 4.1 & 57.4 \\
\bottomrule
\end{tabular}
\label{tab:pos}
\begin{tablenotes}
\item[]Note.-- The quoted XRT positional uncertainties refer to a $90\%$ confidence level. The offset from the \asca\ position is listed in the last column. The two sources that are marked by a dagger have more accurate \chan\ localisations that were recently published and are consistent with our \swift\ positions: AX J1754.2--2754 \citep[][]{bassa08} and AX J1738.4--2902 \citep[][]{evans2010}.
\end{tablenotes}
\end{center}
\end{threeparttable}
\end{table*}

In addition to the 11 firm detections, we assign tentative detections to another 5 sources. For three of these (AX J1720.8--3710, AX J1739.5--2730 and AX J1856.7+0220) we find a clear excess of photons visible by eye within the $2.5$-$\sigma$ \asca\ error circle. The detection significances correspond to $\sim2.5\sigma$ in the full 0.3--10 keV energy band, i.e., just below our chosen threshold. The former two sources had relatively short exposure times ($\lesssim 2$~ks), which is possibly the reason for their low detection significance. 
For two others (AX J1717.2--3718 and AX J1846.1--0239) we detect weak X-ray sources just outside our $90''$ search region at an offset of $1.8'$ from the \asca\ positions (corresponding to a deviation of $3\sigma$). \citet{sugizaki01} indicate that AX J1717.2--3718 was near the edge of the FOV so that the positional accuracy was reduced compared to the mean 90\% confidence error of $1'$ (although no estimate of the positional uncertainty is given). We therefore consider it likely that we indeed detect the XRT counterpart of AX J1717.2--3718 and tentatively also identify the object in the FOV of AX J1846.1--0239 as being related to the \asca\ source. 
We have included the XRT positions of these additional 5 sources in Table~\ref{tab:pos} and the results from spectral analysis are given in Table~\ref{tab:spec}. 

Only few counts were collected for some sources (e.g., AX J1719.3--3703, AX J1725.8--3533, AX J1846.0--0231 and our 5 tentative detections all have $N_{\mathrm{counts}}<35$). Given the very limited statistics, no reliable conclusions on the nature of these sources can be drawn based on their X-ray spectral properties alone. There are two sources with a reasonable number of detected counts ($N_{\mathrm{counts}}>80$) that clearly have different spectral properties than the majority of sources in our sample (Table~\ref{tab:spec}). AX J1738.4--2902 stands out by displaying a considerably less absorbed and softer X-ray spectrum. As we discuss in Sect.~\ref{subsubsec:ms_stars}, the association with an optical/infrared catalogued object makes it likely that this source is a pre-main sequence star. \polar\ appears to have a considerably less absorbed and much harder ($\Gamma\sim0$) spectrum than most sources (see Fig.~\ref{fig:wd}). Based on its spectral properties, we tentatively identify this source as a magnetically accreting white dwarf (see Sect.~\ref{subsubsec:compactobjects}). 

The two brightest sources in our sample, \ucxb\ and \porb, happened to have relatively long exposure times (see Table~\ref{tab:obs}). As a result, we obtained good quality spectra that allowed for an in-depth analysis. This also enabled us to search for any possible X-ray variability on a time scale of days--months. We discuss these two sources separately below.

\begin{table*}
\begin{threeparttable}[t]
\begin{center}
\caption[]{{Count rates, spectral parameters and fluxes for detected sources.}}
\begin{tabular}{l l c c c c c c c c c c c}
\toprule
$\#$ & Name & $N_{\mathrm{counts}}$ & Exp. rate & Obs. rate & Ratio & $N_H^{\mathrm{Gal}}$ & $N_H$ & $\Gamma$ & C-stat (dof) &$F_{\mathrm{X, abs}}$ & $F_{\mathrm{X, unabs}}$ & $L_{\mathrm{X}}$ \\
&  &  & \multicolumn{2}{c}{($10^{-3}~\cnts$)} &  & \multicolumn{2}{c}{($10^{22}~\nh$)} &  &  & \multicolumn{2}{c}{($10^{-12}~\flux$)} & \\
\midrule
\multicolumn{13}{l}{{\bf Firm \swift/XRT detections}} \\ 
2 & AX J1504.6--5824 & 43 & 7.3 & 8.1 & 0.9 & 1.7 & 2.3$^{+3.0}_{-2.1}$ & 1.7$\pm1.4$ & 30.4 (30) & 0.7$^{+0.3}_{-0.7}$ & 1.3 & 1.0 \\
5 & AX J1538.3--5541 & 8622 & 195 & 108 & 1.8 & 1.9 & 7.7$\pm0.5$ & 2.3$\pm$0.1 & 730.6 (745) & 10.3$\pm0.4$ & 56 & 42.9 \\
7 & \polar & 80 & 17.0 & 17.0 & 1.0 & 1.8 &$\le$0.32 & -0.1$\pm0.4$ & 52.2 (58) & 3.1$^{+1.2}_{-3.1}$ & 3.1 & 2.4 \\
8 & AX J1651.0--4403 & 93 & 20.4 & 19 & 1.1 & 1.6 & 0.8$\pm0.6$ & 1.5$\pm0.6$ & 62.8 (72) & 1.3$^{+0.3}_{-1.3}$ & 1.9 & 1.5 \\
12 & AX J1719.3--3703 & 30 & 6.1 & 7.2 & 0.9 & 1.4 & 2.3$^{+4.4}_{-2.0}$ & 0.0$^{+1.3}_{-1.6}$ & 37.6 (25) & 1.2$^{+0.5}_{-1.2}$ & 1.3 & 1.0 \\
14 & AX J1725.8--3533 & 22 & 6.7 & 4.4 & 1.7 & 1.4 & 6.9$^{+11.4}_{-5.9}$ & 1.5$^{+2.6}_{-1.4}$ & 22.5 (17) & 0.6$\pm0.4$ & 1.3 & 1.0 \\
16 & AX J1738.4--2902 & 152 & 22.5 & 31.0 & 0.7 & 0.8 & 0.3$\pm0.2$ & 3.9$\pm0.5$&114.4 (85) & 0.7$\pm0.3$ & 5.1 & 3.9 \\
18 & AX J1742.6--3022 & 72 & 25.2 & 15.0 & 1.7 & 1.2 & 11.0$^{+8.6}_{-4.3}$ & 1.8$\pm2.6$ & 50.2 (56) & 2.0$^{+0.4}_{-2.0}$  & 7.1 & 5.4 \\
19 & AX J1742.6--2901 & 52 & 19.0 & 8.4 & 2.4 & 1.1 & 1.6$\pm0.8$  & 2.9$\pm1.1$ & 40.4 (48) & 0.4$\pm0.3$  & 2.4 & 1.8 \\
22 & AX J1754.2--2754 & 1549 & 177 & 95.0 & 1.9 & 0.9 & 2.1$\pm0.3$ & 2.8$\pm$0.2 &369.7 (458) &5.3$\pm0.5$ & 33   & 33.4 \\
29 & AX J1846.0--0231 & 26 & 2.8 & 2.1 & 1.4 & 1.8 &4.7$\pm4.5$  & 3.2$\pm3.0$ & 7.0 (10) & 0.1$\pm0.1$ & 2.7 & 2.1 \\
\midrule
\multicolumn{13}{l}{{\bf Tentative \swift/XRT detections}} \\ 
11 & AX J1717.2--3718 & 24 & 25.0 & 14.0 & 1.8 & 1.3 & $3.7^{+14.9}_{-3.7}$ & $1.2^{+2.4}_{-1.7}$ & 24.2 (22) & 5.1 & 5.7 & 4.4 \\
13 & AX J1720.8--3710 & 14 & 4.4 & 9.0 & 0.5 & 1.4 & $\le0.34$ & $1.5^{+1.7}_{-0.8}$ & 24.3 (11) & 0.7 & 0.7 & 0.5 \\
17 & AX J1739.5--2730 & 13 & 28.4 & 7.3 & 3.9 & 0.5 & $0.1^{+1.0}_{-0.1}$ & $0.4^{+1.6}_{-1.0}$ & 15.2 (11) & 1.2 & 1.2 & 0.9 \\
30 & AX J1846.1--0239 & 34 & 2.7 & 2.6 & 1.1 & 1.8 & $3.1^{+5.6}_{-3.1}$ & $0.7\pm2.0$ & 17.6 (24) & 0.4 & 0.6 & 0.5 \\
32 & AX J1856.7+0220 & 14 & 4.6 & 2.6 & 1.8 & 1.5 & $3.0^{+7.3}_{-3.0}$ & $0.2^{+2.3}_{-2.2}$ & 7.7 (11) & 0.5 & 0.6 & 0.5 \\
\bottomrule
\end{tabular}
\label{tab:spec}
\begin{tablenotes}
\item[]Note. -- All count rates, fluxes and luminosities quoted in this table refer to the 0.3--10 keV energy band. The errors on the spectral parameters refer to 90\% confidence levels. For each source we list the total number of detected counts ($N_{\mathrm{counts}}$), the expected (exp.) and observed (obs.) XRT count rates (see text), as well as the the ratio of these two (expected/observed). We give both the Galactic \citep[$N_H^{\mathrm{Gal}}$; from][]{kalberla2005} and observed ($N_H$) hydrogen column density for each source. The X-ray luminosity is expressed in units of $10^{34}~\lum$ and calculated from the unabsorbed flux ($F_{\mathrm{X, unabs}}$) by assuming a distance of $D=8$~kpc for all sources, except AX J1754.2--2754 for which we adopt D=9.2 kpc \citep[][]{chelovekov07}. 
\end{tablenotes}
\end{center}
\end{threeparttable}
\end{table*}

\subsubsection{\ucxb}\label{subsubsec:ucxb}
\ucxb\ was discovered during \asca\ observations performed on 1999 October 2 \citep[][]{sakano02}. At the time, the source could not be associated with any of the known X-ray sources from the \einstein\ or \rosat\ catalogues and remained unclassified. \inte\ detected a thermonuclear X-ray burst from \ucxb\ in 2005  \citep[which was reported in 2007 after we submitted our target list for this program;][]{chelovekov07}. This unambiguously identified the previously unclassified \asca\ source as a neutron star accreting from a (sub-) solar mass companion star: a so-called low-mass X-ray binary (LMXB). This also allowed for a distance estimate of $D\lesssim9.2$~kpc \citep[][]{chelovekov07}. 

Since its discovery with \asca\ in 1999, \ucxb\ was found active with \swift, \chan\ and \inte\ on several occasions, displaying a typical 0.5--10 keV luminosity of $L_X \sim 10^{35}~(D/9.2~\mathrm{kpc})^2~\lum$ \citep[][]{delsanto07,krivonos07,bassa08,jonker08}. In 2008 May, however, the source could not be detected with \chan. This implied that its 0.5--10 keV luminosity was $L_X \lesssim 5\times 10^{32}~(D/9.2~\mathrm{kpc})^2~\lum$ and suggested a transient nature \citep[][]{bassa08}. Follow-up observations performed in 2008 July--August detected the source back at its outburst level \citep[][]{jonker08}. 

\begin{figure*}
 \begin{center}
 \includegraphics[width=9cm]{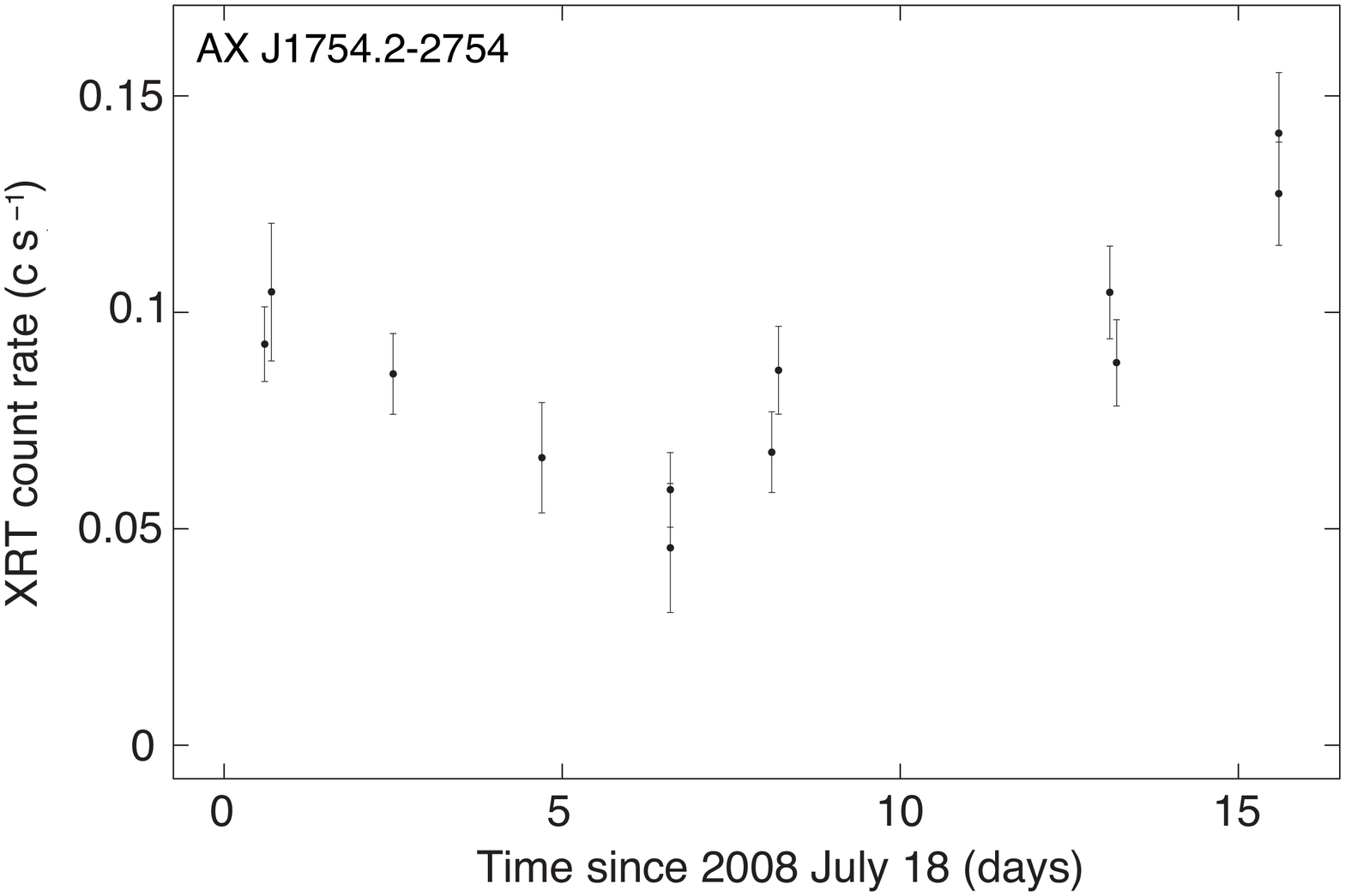}
          \hspace{0.5cm}
\includegraphics[width=8.5cm]{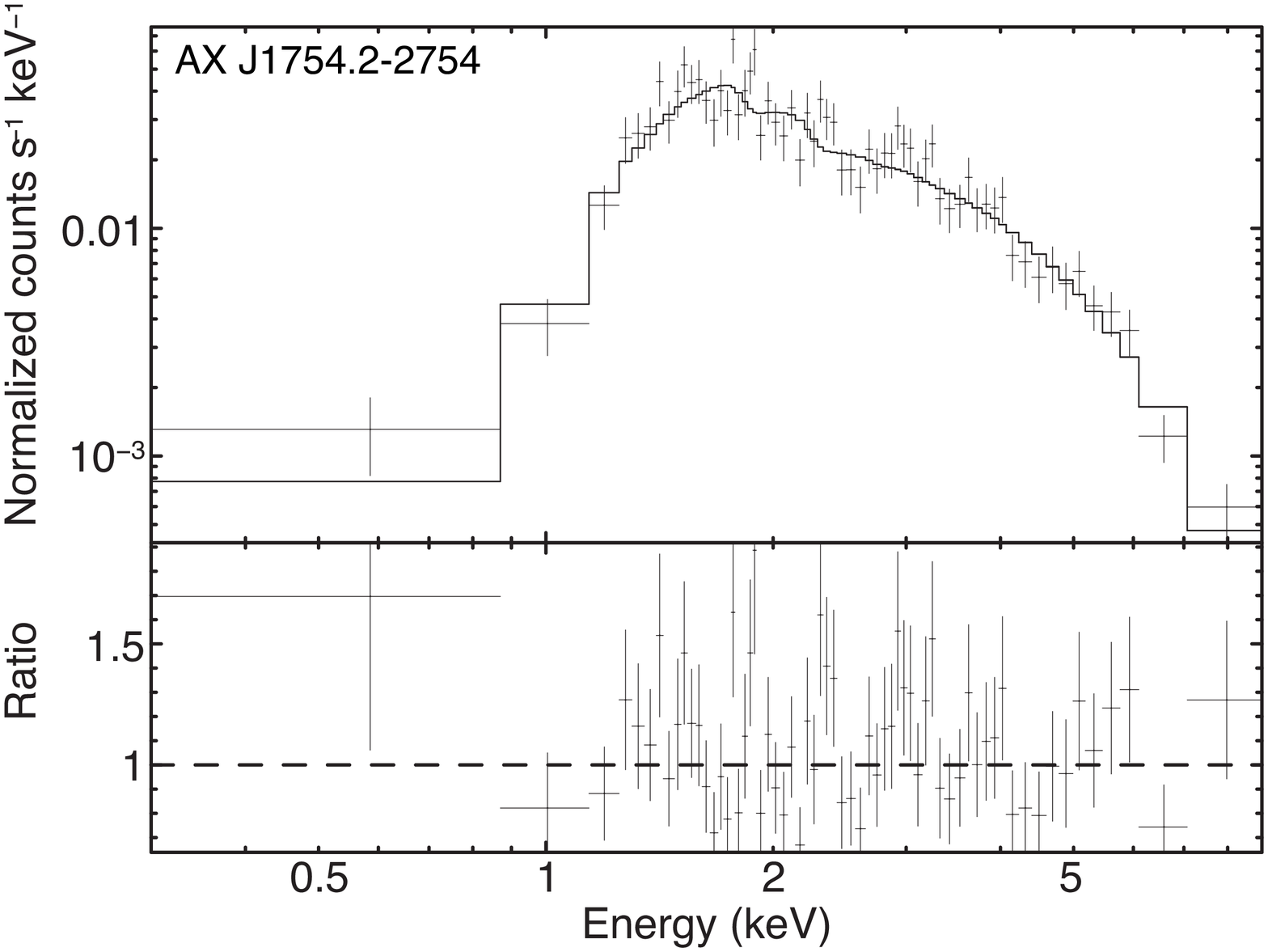}
    \end{center}
\caption[]{\swift/XRT results for the confirmed neutron star LMXB \ucxb. Left: The X-ray light curve observed in 2008 July--August (binned per satellite orbit). Right: The average X-ray spectrum obtained by summing all available data (2007 July--2008 August). The solid line represents the best-fit result to an absorbed powerlaw model and the bottom plot indicates the data to model ratio of the fit. 
}
 \label{fig:ucxb}
\end{figure*}

All \swift/XRT observations of \ucxb\ are listed in Table~\ref{tab:obs}. The source was observed twice in 2007 July \citep[][]{delsanto07_ascabron} and a series of pointings were performed in 2008 July--August following the \chan\ non-detection \citep[][]{jonker08}. Figure~\ref{fig:ucxb} displays the \swift/XRT count rate light curve of \ucxb\ obtained during this epoch. It shows that the source remained active with an intensity that varied by a factor of $\sim3$ on a time scale of days. We summed all available XRT data (amounting to 16.3~ks of exposure) to obtain an average X-ray spectrum of \ucxb\ (Fig.~\ref{fig:ucxb}). We grouped the spectral data to contain at least 20 photons per bin and performed fits using chi-squared ($\chi^2$) statistics.

A fit to an absorbed powerlaw model suggests that the source is strongly absorbed [$N_H = (2.1\pm0.3) \times10^{22}~\nh$] and has a rather soft spectral index of $\Gamma =2.7\pm0.2$. This yields a reduced chi-squared value of $\chi_{\nu}^2=1.11$ for 61 degrees of freedom (dof). The resulting average 0.3--10 keV luminosity is $L_X \sim 3\times10^{35}~(D/9.2~\mathrm{kpc})^2~\lum$, or $L_X \sim 6\times10^{34}~(D/9.2~\mathrm{kpc})^2~\lum$ in the often quoted 2--10 keV band.\footnote{The 2--10 keV energy band is often used in literature to characterise and classify X-ray sources \citep[e.g.,][]{wijnands06}. When appropriate, we therefore give the 2--10 keV luminosities in addition to the 0.3--10 keV intensities that are used throughout this work.}  These results are consistent with those obtained using C-statistics (see Table~\ref{tab:spec}). 

The high spectral index implied by the simple absorbed powerlaw fit suggests that the spectrum possibly has a thermal component. However, adding a soft spectral component (e.g., BBODYRAD or DISKBB) to the powerlaw model does not improve the fit ($\chi_{\nu}^2=1.11$ for 59 dof). A single absorbed blackbody model (BBODYRAD) cannot acceptably describe the XRT spectral data ($\chi_{\nu}^2=1.71$ for 61 dof).

\subsubsection{\porb}\label{subsubsec:porb}
\porb\ was discovered with \asca\ on 1998 March 11 \citep[][]{sugizaki01}. In our \swift\ program, the source region was observed with the XRT during four short exposures obtained in 2007 January--March and it was monitored during a longer series of pointings in the epoch of 2007 August--September (see Table~\ref{tab:obs}). The total XRT exposure time on this target amounts to 101.9 ks. The distance towards this object is unknown and we assume a fiducial value of $D=8$~kpc.

\begin{figure*}
 \begin{center}
 	\includegraphics[width=9cm]{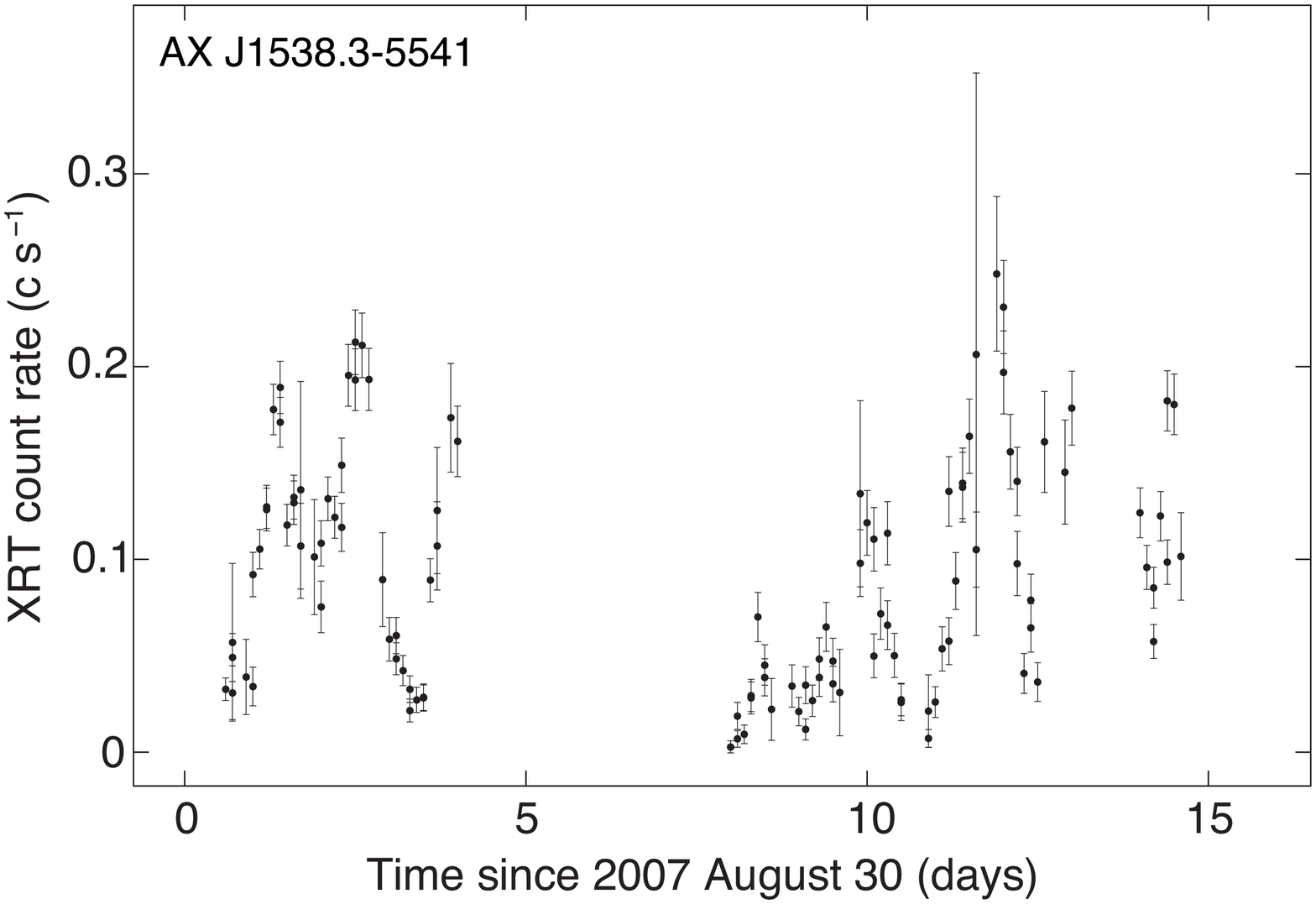}
         \hspace{0.5cm}
 \includegraphics[width=8.5cm]{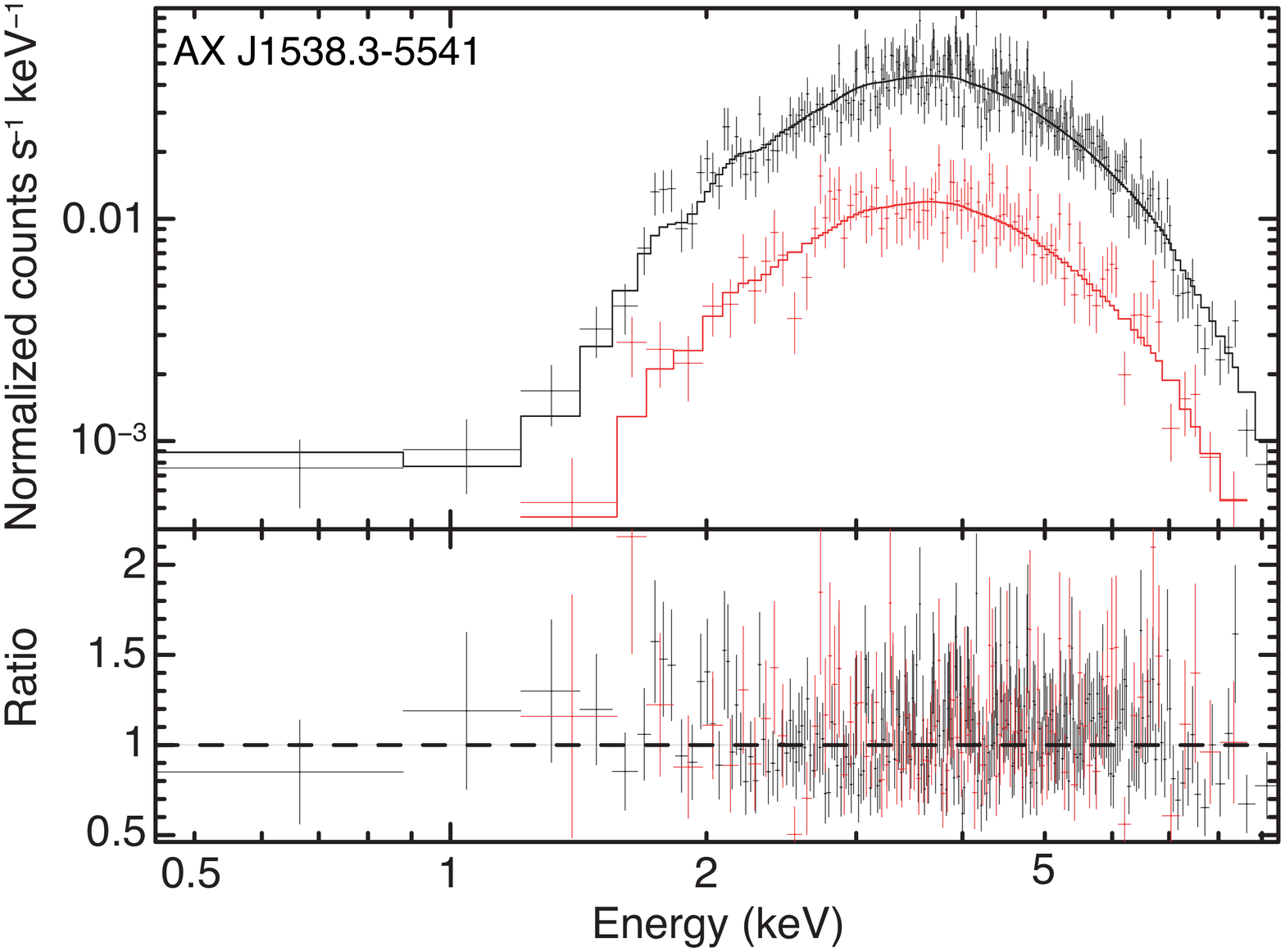}
    \end{center}
\caption[]{\swift/XRT results for \porb. Left: X-ray count rate light curve of the epoch of 2007 August--September (binned per satellite orbit). Right: Comparison of the X-ray spectra at high (average of $0.16 ~\cnts$; top, black) and low (average of $0.04~\cnts$; bottom, red) count rates, using all available \swift/XRT observations of the source. The solid lines indicate the best-fit results to an absorbed powerlaw model. The lower panel indicates the data to model ratio of the fits. 
}
 \label{fig:porb}
\end{figure*}

Figure~\ref{fig:porb} shows the X-ray light curve of \porb\ obtained during the intense monitoring of 2007 August--September. It is clear from this plot that the X-ray emission is strongly variable, changing by a factor of $\sim10$ on a timescale of days. The source was detected at a maximum and minimum XRT count rate of $0.26$ and $7.0\times10^{-3}~\cnts$, respectively. Using a count rate to flux conversion factor inferred from fitting the average XRT spectrum (see below), we find that the source reached a peak 0.3--10 keV luminosity of $L_X \sim 1.5\times10^{36}~(D/8~\mathrm{kpc})^2~\lum$ during our observations, while its lowest intensity was $L_X \sim 3.8\times10^{34}~(D/8~\mathrm{kpc})^2~\lum$. In the 2--10 keV energy band the extrema translate into luminosities of $L_X \sim 4.6\times10^{35}$ and $L_X \sim 1.2\times10^{34}~(D/8~\mathrm{kpc})^2~\lum$.

Preliminary analysis of a sample of XRT observations revealed indications of a (non-sinusoidal) modulation in the X-ray light curve of $1.23 \pm 0.05$ days \citep[][]{kennea07}. To search for periodic variations in \porb, we created a Lomb Scargle periodogram from all the XRT data shown in the light curve in Fig.~\ref{fig:porb}, which spans 14 days of observations for a total exposure time of 80.7~ks. We find no periodic variability and hence our analysis does not confirm the possible period of 1.23 days that was found when analysing a subset of these data.

We summed all observations and examined the average X-ray spectrum of \porb. The spectral data was grouped to contain a minimum of 20 counts per bin and fitted using $\chi^2$-statistics. A simple absorbed powerlaw model can adequately describe the data ($\chi_{\nu}^2=1.05$ for 225 dof) and suggests that the spectrum is highly absorbed [$N_H = (8.0\pm0.5)\times10^{22}~\nh$] and relatively soft ($\Gamma=2.3\pm0.1$). The average unabsorbed 0.3--10 keV flux of the powerlaw fit translates into a luminosity of $L_X \sim 4.6\times10^{35}~(D/8~\mathrm{kpc})^2~\lum$ or $L_X \sim 1.5\times10^{35}~(D/8~\mathrm{kpc})^2~\lum$ in the 2--10 keV energy range. The average X-ray flux inferred from our XRT observations is a factor of $\sim2$ lower than that detected with \asca\ in 1998 \citep[][]{sugizaki01}. An absorbed blackbody model (BBODYRAD) can describe the data equally well ($\chi_{\nu}^2=1.09$ for 225 dof) and yields $N_H = (4.3\pm0.3)\times10^{22}~\nh$, a temperature of $kT_{bb}=1.3\pm0.1$~keV and an emitting radius of $R_{bb}=0.54\pm0.04~(D/8~\mathrm{kpc})$~km. 

We investigated whether the strong variability in X-ray intensity was accompanied by changes in the X-ray spectrum. To compare the spectral shape at high and low count rates, we extracted spectra of \porb\ using all the available XRT observations (Table~\ref{tab:obs}) and making count rate cuts of $\ge 0.12~\cnts$ and $\le 0.09~\cnts$ (see the light curve in Fig.~\ref{fig:porb}). This resulted in exposure times of 42.3 and 36.0~ks, and averaged count rates of 0.16 and $0.04~\cnts$ for the high and low count rate regimes, respectively. Both spectra were extracted in an identical fashion using a 30 pixel radius source region centred at the XRT-measured position (Table~\ref{tab:pos}), and a nearby source-free region of radius 100 pixels as a background reference. The high count rate spectrum was grouped such that a minimum of 20 counts fall in each bin and we required a 15-counts minimum per bin in the low count rate spectrum.

\begin{table*}
\begin{threeparttable}[t]
\begin{center}
\caption[]{{Upper limits on the count rates and fluxes of the sources that were not detected.}}
\begin{tabular}{l l c c c c c c c}
\toprule
\# & Name & Exp. rate & Obs. rate & Ratio & $N_H^{\mathrm{Gal}}$ & $F_{\mathrm{X, abs}}$ & $F_{\mathrm{X, unabs}}$ & $L_{\mathrm{X}}$  \\
&  &  \multicolumn{2}{c}{($10^{-3}~\cnts$)} &  & ($10^{22}~\nh$) & \multicolumn{2}{c}{($10^{-12}~\flux$)} & ($10^{33}~\lum$) \\ 
\midrule 
1 & AX J1457.5--5901 & 8.34 & $<2.47$ & $>3.38$ & 1.7 & $<0.2$ & $<0.4$ & $<3$ \\
3 & AX J1510.0--5823 & 5.60 & $<2.78$ & $>2.00$ & 1.7 & $<0.2$ & $<0.4$  &  $<3$ \\
4 & AX J1537.8--5556 & 6.40 & $<2.11$ & $>3.04$ & 1.9 &  $<0.2$ & $<0.3$  &  $<2$ \\
6 & AX J1545.9--5443 & 2.91 & $<2.19$ & $>1.33$ & 1.7 &  $<0.2$ &  $<0.3$ & $<2$  \\
9 & AX J1657.3--4321 & 2.66 & $<2.73$ & $>0.97$ & 1.6 & $<0.2$  & $<0.4$  & $<3$  \\
10 & AX J1659.8--4209 & 3.30 & $<4.72$ & $>0.70$ & 1.6 &  $<0.3$ &  $<0.7$ & $<5$  \\
15 & AX J1734.5--2915 & 7.76 & $<2.44$ & $>3.18$ & 0.6 &  $<0.1$ & $<0.2$  & $<2$  \\
20 & AX J1751.1--2748 & 27.5 & $<3.90$ & $>7.04$ & 1.2 &  $<0.3$ &  $<0.5$ & $<4$  \\
21 & AX J1753.5--2745 & 4.34 & $<3.06$ & $>1.42$ & 1.0 &  $<0.2$ &  $<0.3$ & $<2$  \\
23 & AX J1758.0--2812 & 4.21 & $<5.10$ & $>0.83$ & 0.5 &  $<0.3$ &  $<0.4$ & $<3$  \\
24 & AX J1832.5--0916 & 2.58 & $<3.65$ & $>0.71$ & 1.6 &  $<0.3$ &  $<0.5$ & $<4$  \\
25 & AX J1833.9--0822 & 18.7 & $<3.31$ & $>5.66$ & 1.7 &  $<0.2$ & $<0.5$  & $<4$  \\
26 & AX J1834.6--0801 & 2.49 & $<2.26$ & $>1.10$ & 1.8 & $<0.2$  & $<0.3$  & $<2$  \\
27 & AX J1835.1--0806 & 3.02 & $<2.22$ & $>1.36$ & 1.8 & $<0.2$  &  $<0.3$ &  $<2$ \\
28 & AX J1836.3--0647 & 27.6 & $<2.82$ & $>9.80$ & 1.6 & $<0.2$  & $<0.4$  & $<3$  \\
31 & AX J1847.6--0219 & 5.21 & $<1.95$ & $>2.67$ & 1.8 &  $<0.1$ &  $<0.3$ & $<2$  \\
33 & AX J1856.8+0245 & 1.91 & $<1.36$ & $>1.40$ & 1.5 & $<0.1$  & $<0.2$  & $<2$  \\
34 & AX J1857.3+0247 & 2.29 & $<1.28$ & $>1.79$ & 1.5 &  $<0.1$ & $<0.2$  & $<2$  \\
35 & AX J1900.1+0427 & 2.80 & $<2.17$ & $>1.29$ & 1.7 &  $<0.1$ & $<0.3$  & $<2$  \\
\bottomrule
\end{tabular}
\label{tab:nondet}
\begin{tablenotes}
\item[]Note. -- All quoted count rates, fluxes and luminosities are given in the 0.3--10 keV energy range. We list both the expected (exp.) and observed (obs.) XRT count rates (see text), as well as the ratio of the two (expected/observed). The listed $3\sigma$ count rate upper limits were translated into fluxes using \textsc{pimms} by assuming a fiducial absorbed powerlaw model with a spectral index of $\Gamma=1.9$ and a hydrogen column density corresponding to the listed Galactic values from \citet{kalberla2005}. The luminosities were calculated from the unabsorbed flux by adopting a distance of 8 kpc for all sources.
\end{tablenotes}
\end{center}
\end{threeparttable}
\end{table*}

Figure~\ref{fig:porb} shows the two spectra and the data to model ratios for an absorbed power law fit. Fitting to the high and low count rate regimes independently, we find they can be described by the same spectral model within the errors of $\Gamma = 1.8 \pm 0.1$, $2.0 \pm 0.3$ and $N_H = (6.8\pm0.5) \times10^{22}$, $(7\pm1)\times10^{22}~\nh$ for the high and low count rates, respectively. These agree well (within 90\% errors) with the spectrum measured by \asca\ \citep[][]{sugizaki01}. While the uncertainties are quite large on the low count rate spectrum, we can conclude that strong spectral evolution is not occurring in these data, despite a factor $\sim4$ variability in count rate (see Fig.~\ref{fig:porb}). We also investigated the relation between the 0.3--10 keV count rate and the spectral hardness, which we defined as the ratio of counts in the 3--10 and 0.3--3 keV energy bands for the present purpose. Again, we found no indications of a change in spectral shape with changing intensity. Based on the observed X-ray properties we speculate that \porb\ may be an X-ray binary, which we discuss in more detail in Sect.~\ref{subsubsec:compactobjects}.

\subsection{X-ray non-detections}\label{subsec:nondet}

With 11 firm and 5 tentative detections, our \swift/XRT program detected 16 out of 35 observed sources (i.e., $46\%$). We do not detect XRT counterparts of 19 sources in our sample (i.e., $54\%$). We list the inferred $3\sigma$ count rate upper limits from our \swift/XRT observations in Table~\ref{tab:nondet}, together with the expected XRT count rates (see Sect.~\ref{obs_ana}) and the ratio of the two. Corresponding estimates of the upper limits on the 0.3--10 keV fluxes and luminosities are also given.

We note that in three cases we found X-ray sources in the FOV that were detected at such a large offset that an association with the \asca\ source seems unlikely. This concerns AX J1833.9--0822, AX J1847.6--0219 and AX J1857.3+0247, for which we detect faint X-ray sources at offsets of $5.2'$, $5.7'$ and $3.5'$, respectively (corresponding to $8.5\sigma$, $9.3\sigma$ and $5.7\sigma$ deviations from the \asca\ positions). We assign these three sources \swift\ names and list their positions, count rates and possible associations to catalogued objects in Table~\ref{tab:other}.

\begin{table*}
\begin{threeparttable}[t]
\begin{center}
\caption[]{{Positions (J2000) and errors for additional sources detected during our observations.}}
\begin{tabular}{l c c c c l}
\toprule
Name & R.A. & Dec. & Error & Rate & Possible association \\
& (h m s) & $(^{\circ}~'~'')$ & (\arcsec) & ($\cnts$) & \\
\midrule
Swift J183345.2--081831 & 18 33 45.24  & -08 18 31.0 & 2.8 & $1.0\times10^{-2}$ & Star BD-08 4632 at $1.7''$ from the XRT position \\
Swift J184803.6--021805 & 18 48 03.61 & -02 18 05.9 & 3.9 & $8.2\times10^{-3}$ & {\it NOMAD/2MASS} object at $2.2''/1.8''$ from the XRT position\\
Swift J185712.6+025010 & 18 57 12.61 & +02 50 10.7 & 5.2 & $3.6\times10^{-3}$ & \ldots  \\
\bottomrule
\end{tabular}
\label{tab:other}
\begin{tablenotes}
\item[]Note.-- The quoted XRT positional uncertainties refer to a $90\%$ confidence level and the count rates are for the 0.3--10 keV energy band.
\end{tablenotes}
\end{center}
\end{threeparttable}
\end{table*}

\subsection{Possible associations at other wavelengths}\label{subsec:oir}
Searching within the 90\% confidence uncertainties of the XRT positions, we detected 4 possible counterparts in simultaneously obtained \swift/UVOT observations. The offset from the XRT positions and the inferred UV/optical magnitudes are given in Table~\ref{tab:oir}. In this table, we also list possible associations with catalogued astronomical objects at optical (6 candidates) and infrared wavelengths (7 candidates). In total, we found possible UV/optical/infrared counterparts for 8 of the XRT-detected sources in our sample. Three of these have positions consistent with \rosat\ objects and one may also be related to a radio source (see below). 

We note that the sources for which we find possible counterparts have XRT positional uncertainties of $\ge 2.2''$ (see Table~\ref{tab:pos}), which makes the probability of chance detections relatively high in these crowded fields. For instance, a catalogue search in a $1'$ circular region around the XRT position of AX J1504.6--5824 returns 101 {\it NOMAD}-objects (and 97 {\it 2MASS}-objects). This implies that the chance of detecting an uncorrelated source within the $4.1''$ XRT uncertainty of AX J1504.6--5824 is about $50\%$. We find similar probabilities of optical/infrared chance detections for the other sources listed in Table~\ref{tab:oir}. Further evidence leaning to a considerable number of chance detections is illustrated by the fact that the offset with the possible counterparts is typically close to our 90\% XRT positional uncertainties (see Tables~\ref{tab:pos} and~\ref{tab:oir}). 

A preliminary determination of the \swift/XRT position of \porb\ yielded a 90\% confidence uncertainty of $3.6''$, which included two {\it 2MASS}-objects that both have a magnitude of $J\sim14$ \citep[][]{kennea07}. By using additional \swift\ observations, we have improved the positional uncertainty of \porb\ to $1.9''$ (see Table~\ref{tab:pos}). Both sources from the {\it 2MASS} catalogue lie outside this 90\% confidence error circle and are thus not likely to be related to the X-ray source. This also demonstrates the high probability of a chance detection, even within positional uncertainties as small as a few arcseconds. We highlight some of the sources with potential counterparts below.

\subsubsection{AX J1738.4--2902 and AX J1846.1--0239}
The improved XRT positions of two \asca\ sources have relatively small offsets from potential optical/infrared counterparts, which makes an association likely. Our XRT position of AX J1738.4--2902 is only $0.2''$ away from the \rosat\ source 1RXS J173826.7--290140 \citep[][]{sidoli01}, which has an accurate \chan\ localisation \citep[][]{evans2010,jonker2011} and is associated with the infrared object 2MASS J17382620--2901494. This object is classified as a pre-main sequence star of spectral type K0V \citep[][]{skiff2009}. The small offset between our XRT position and this object, combined with the low absorption column density and soft X-ray spectrum inferred from our XRT observations (Table~\ref{tab:spec}), make it likely that this is the object detected during our observations. We therefore classify AX J1738.4--2902 as a K0V main sequence star (see also Sect.~\ref{subsubsec:ms_stars}).

For AX J1846.1--0239 we find three {\it 2MASS} objects located within the $4.1''$ XRT positional uncertainty, of which the most distant one appears to coincide with an optical source (see Table~\ref{tab:oir}). One of the infrared objects is located only $0.8''$ from our XRT position and has particularly red colours (see Sect.~\ref{subsubsec:ms_stars}). The close proximity suggests that an association might be likely, although the fact that we find three possible infrared counterparts within our X-ray positional uncertainty underlines the high probability of a chance detection. We collected only a small number of photons for AX J1846.1--0239 in our XRT spectrum, so we cannot further speculate on the nature of this object and the likelihood of an infrared association based on its X-ray spectral properties (see Table~\ref{tab:spec}).

\subsubsection{AX J1742.6--2901 and AX J1720.8--3710}
Apart from AX J1738.4--2902, there are two more sources in our sample that are potentially related to \rosat\ objects. Our XRT position of AX J1742.6--2901 coincides with 2RXP J174241.8--290215 and lies $2.5''$ from a radio source listed in the 6- and 20-cm survey of the Galactic centre \citep[][]{law2008}. This is well within our $3.4''$ XRT positional uncertainty (see Table~\ref{tab:pos}). The region around AX J1742.6--2901 was covered with \chan\ during a monitoring campaign of 1.2 square degrees around the Galactic centre. A faint X-ray source was detected at a position consistent with our XRT localisation. Due to the fact that the source was located at large offset angles from the aimpoint of the \chan\ observations the derived position is less accurate than inferred from our \swift/XRT observations (Degenaar et al. in preparation).

AX J1720.8--3710, is $3.2''$ away from the (unclassified) X-ray source 1RXS 172051.9--371033 and might also be related to an infrared object whose colours match that of a main sequence star (see Sect.~\ref{subsubsec:ms_stars}). Since we infer a small to negligible hydrogen absorption column density from fitting the XRT spectrum (see Table~\ref{tab:spec}), we find it likely that the X-ray source that we identify as AX J1720.8--3710 is related to this (foreground) star.

\subsection{Possible high-energy counterparts}\label{subsec:hard}
Cross-correlation of our XRT positions with the hard X-ray sources from the \swift/BAT 58-month survey\footnote{Listed at: http://swift.gsfc.nasa.gov/docs/swift/results/bs58mon/} yields no matches [Baumgartner et al. submitted to ApJS, see also \citet{tueller09} and the 54-months Palermo BAT survey; \citet{cusumano2010}]. There are thus no hard counterparts for our XRT-detected sources in the BAT survey. Nevertheless, two of the sources from our sample have tentative high-energy counterparts proposed elsewhere. \citet{bouchet2005} find a hard X-ray source in an \inte\ survey of the Galactic central radian that they tentatively associate with AX J1758.0--2818. We do not detect this \asca\ source during our \swift/XRT observations.

\citet{hessels2008} propose that AX J1856.8+0245 could be the X-ray counterpart to the young energetic pulsar PSR J1856+0245, which is likely associated with the TeV gamma-ray source HESS J1857+026. AX J1856.8+0245 is not detected during our program. These authors also mention that the nearby \asca\ source AX J1857.3+0247 might be related to some extended emission of HESS J1857+026. We also targeted AX J1857.3+0247 (see Table~\ref{tab:obs}), but did not detect the source during our XRT observations. The non-detection could be due to the source being extended rather than point-like, since this would lower our sensitivity \citep[see e.g.,][]{cackett06}. However, we argue in Sect.~\ref{subsec:nat_nondetect} that AX J1857.3+0247 potentially concerns a spurious \asca\ detection.

\begin{table*}
\begin{threeparttable}[tb]
\begin{center}
\caption{UV, optical and infrared magnitudes of possible associations to XRT-detected sources. }
\begin{tabular}{l l c c c c c c}
\toprule
 &  & &  & $uvw2$ & $u$ & $b$ & $v$  \\
$\#$ & Name  & Catalogue & Offset ($''$)  & $B$ & $V$ & $R$ & $I$  \\
 &  &  &  & $J$ & $H$ & $K$ & \\
\midrule
2 & AX J1504.6--5824 & {\it UVOT} & 3.9 & \ldots & \ldots & $19.3\pm0.1$ & $17.8\pm0.1$ \\
 &  & {\it NOMAD/USNO-B1}  & 3.7 & 18.46  & 17.22 & 16.41/16.87 & 15.62   \\
 &  & {\it 2MASS} & 3.9 & $13.72 $ & $13.14 $ & $12.77$  & \ldots \\
11 & AX J1717.2--3718 & {\it 2MASS} & 2.2 & $17.62$ & $15.70 $ & $13.52 $ & \ldots \\
13 & AX J1720.8--3710$^{+}$ & {\it UVOT} & 4.0 & $15.02\pm0.02$ & \ldots & \ldots &  \ldots  \\
 &  & {\it NOMAD/USNO-B1} & 3.9 & $12.14/11.97$ & $11.81 $ & $11.64/11.62$ & $11.47 $  \\
 &  &  {\it 2MASS} & 3.8 & $9.90 $ & $9.55 $ & $9.44 $ & \ldots \\
16 & AX J1738.4--2902$^{+}$ & {\it UVOT} & 0.2 & $14.82\pm0.01$ & \ldots & \ldots & \ldots   \\
 &  & {\it NOMAD/USNO-B1} & 0.2 & 12.21/11.67 & 11.20 & 10.70/10.64 & 10.22  \\
 &  & {\it 2MASS}& 0.4 & $9.39 $ & $8.92 $ & $8.82 $ & \ldots \\
17 & AX J1739.5--2730  & {\it NOMAD/USNO-B1} & 3.3 & 19.43 & \ldots & 17.34/17.98  &  \ldots  \\
19 & AX J1742.6--2901$^{+}$ & {\it UVOT} & 2.8 & \ldots &  $18.55\pm0.06$ & \ldots &  \ldots \\
 &  & {\it NOMAD/USNO-B1} & 2.4& 17.96/16.93&14.25& 12.66/12.54& $10.28$  \\
 &  & {\it 2MASS}& 2.5& 8.11 & 7.27 & 6.73 &  \ldots  \\
29 & AX J1846.0--0231 &{\it 2MASS} & 3.7 & 14.52 & 12.84 & 12.20 &  \ldots   \\
30 & AX J1846.1--0239 &{\it 2MASS}  & 0.8 & 14.96 & 12.84 & 10.44 & \ldots    \\
 &  &{\it 2MASS}  & 3.6 & 13.45 & 10.94 & 10.63 &  \ldots   \\
 &  & {\it NOMAD/USNO-B1} & 4.1 & 21.01/20.00 & \ldots & 18.98/18.15 & 17.26 \\
 &  &{\it 2MASS}  & 3.9 & 15.27 & 14.42 & 12.53 &  \ldots \\
\bottomrule
\end{tabular}
\label{tab:oir}
\begin{tablenotes}
\item[]Note. -- We searched for possible counterparts within the 90\% XRT error radii. The fourth column represents the offset between the centres of the possible counterparts and the XRT positions. Quoted UV/optical magnitudes were taken either from \swift/UVOT data ($uvw2$, $u,b,v$), the Naval Observatory Merged Astrometric Dataset \citep[{\it NOMAD};][$V$-band]{zacharias2004} or the USNO-B1.0 catalogue \citep[][$BRI$]{monet2003}, while the infrared magnitudes come from the {\it 2MASS} all-sky catalogue of point sources \citep[][$JHK$]{cutri2003}. All catalogue magnitudes are given without errors and rounded to two decimal places. Sources marked by a plus-symbol may be associated with \rosat\ objects (see text and Table~\ref{tab:class}) and the position of one of these is also consistent with a radio source (AX J1742.6--2901; see text).
\end{tablenotes}
\end{center}
\end{threeparttable}
\end{table*}

\section{Discussion}\label{sec:discuss}
We carried out \swift\ follow-up observations of a sample of 35 unclassified faint X-ray sources detected during \asca\ surveys of the Galactic plane and centre between 1993 and 1999 \citep[][]{sugizaki01,sakano02}. A total of 16 sources from our sample were detected with the XRT, which allows us to improve their positional accuracy from $\sim$1$'$ to $\sim$2--4$''$. The mean offset of the improved XRT position from the \asca\ coordinates is $47.9''$ ($0.8'$), which is relatively close to the reported 90\% confidence \asca\ positional uncertainties. 

We investigated the X-ray spectra of the detected sources and studied any possible long-term X-ray variability of all our 35 targets. Our program was conducted between 2006 and 2010, hence $\sim7-17$ years since the \asca\ detections. In general, it is not possible to classify the sources based on their X-ray spectral properties, particularly given the fact that the total number of counts detected during our observations was often low (see Table~\ref{tab:spec}). Using the improved \swift/XRT position information, we therefore searched optical and infrared source catalogues to assess any possible associations. Below we discuss the possible nature of the sample of unclassified \asca\ sources covered by our program. Proposed classifications and possible associations are summarised in Table~\ref{tab:class}.

\subsection{The nature of the \asca\ sources detected with \swift}\label{subsec:nat_detect}

\subsubsection{Accreting compact objects}\label{subsubsec:compactobjects}

{\bf \ucxb} was observed with \swift/XRT on two different epochs in 2007 July and 2008 August--July, during which it is detected at an average 2--10 keV luminosity of $L_X \sim 6\times10^{34}~(D/9.2~\mathrm{kpc})^2~\lum$. This is considerably lower than typically observed for neutron star LMXBs and classifies the source as a sub-luminous X-ray binary \cite[][]{muno05_apj622,wijnands06,degenaar09_gc,campana09}. The \swift/XRT data show that the quiescent phase signalled by \chan\ in 2008 May must have been short \citep[$\lesssim11$~months, see][]{bassa08,jonker08}. Its ability to spend long times accreting at a low X-ray luminosity, combined with its thermonuclear X-ray burst properties and low absolute optical magnitude ($M_i \gtrsim 2$), led to the suggestion that \ucxb\ could be an ultra-compact X-ray binary \citep[][]{bassa08}. Such systems harbour hydrogen-poor donor stars in small orbits of $\lesssim 90$~min \citep[][]{nelson86}. 

The XRT spectrum of \ucxb\ is very soft with a photon index of $\Gamma=2.7$. Such a high spectral index is quite unusual, although other LMXBs accreting at similarly low luminosities have also been found to display very soft X-ray spectra \citep[e.g.,][]{zand05,delsanto07,degenaar09_gc,degenaar2010_burst,armas2011}. An in-depth investigation of high-quality \xmm\ data of the sub-luminous (candidate) LMXB \xte\ revealed that the X-ray spectrum contained a soft (thermal) component in addition to the hard powerlaw emission \citep[][]{armas2011}. In contrast, the XRT spectrum of \ucxb\ can be adequately fit by an absorbed powerlaw alone and does not require the addition of a soft spectral component. \\

\noindent {\bf \porb} is the brightest source in our sample and was detected with the XRT at an average 2--10 keV luminosity of $L_X \sim 2\times10^{35}~(D/8~\mathrm{kpc})^2~\lum$. It has a highly absorbed ($N_H \sim 8 \times 10^{22}~\nh$) X-ray spectrum, that can be described by a simple powerlaw model with an index of $\Gamma=2.3$. The XRT light curve shows that the source intensity varies considerably by a factor of $\sim10$ on a timescale of days. Given its X-ray spectral characteristics, its average intensity and the timescale of the observed variability we tentatively identify \porb\ as an X-ray binary. 

During our program \porb\ spanned a luminosity range of $L_X\sim1\times10^{34}-5\times10^{35}~(D/8~\mathrm{kpc})^2~\lum$. Such large variations in X-ray intensity (a factor $\sim50$) are not uncommon for X-ray binaries and are usually ascribed to considerable variations in the mass-accretion rate onto the compact primary. For example, strong X-ray variability is a characteristic feature of Supergiant Fast X-ray Transients (SFXTs), in which the compact primary accretes matter from the varying wind of a massive ($\gtrsim10~\Msun$) supergiant companion star \citep[e.g.,][]{sidoli08,romano2011}. Similar to what we observe for \porb, these kind of systems are typically highly absorbed. The comparison breaks down, however, when considering the X-ray spectral shape: SFXTs typically have much harder X-ray spectra ($\Gamma \sim 1$) in the luminosity range that we observe for \porb\ \citep[][]{romano2011}. 

Variations of similar magnitude and on a similar timescale as seen for \porb\ have also been observed from LMXBs in which the companion star overflows its Roche lobe and accretion is governed via an accretion disk \citep[e.g.,][]{wijnands2001_1808,linares2008,degenaar2011_1701,fridriksson2011}. Although the average 2--10 keV X-ray luminosity of this source is lower than is typically seen for active LMXBs \citep[e.g.,][]{chen97}, there is a growing group of sub-luminous sources that accrete at similar X-ray intensities to \porb\ \citep[e.g., \ucxb\ discussed above, but see also][]{cornelisse02,muno05_apj622,wijnands06,zand05,zand07,delsanto07,degenaar09_gc,campana09,degenaar2010_burst}. 

Based on its X-ray spectral characteristics we consider it more likely that \porb\ belongs to the class of LMXBs and thus harbors a low-mass ($\lesssim1~\Msun$) companion star as opposed to a more massive donor. Dedicated follow-up observations at longer wavelengths have the potential to rule out a high-mass X-ray binary nature and thus provide a test for our proposed classification.\\

\noindent {\bf \polar} stands out by displaying a much less absorbed ($N_H\lesssim3\times10^{21}~\nh$) and much flatter ($\Gamma \sim 0$) X-ray spectrum than observed for the majority of X-ray sources in our sample (see Table~\ref{tab:spec}). The hard X-ray spectrum of \polar\ (see Fig.~\ref{fig:wd}) resembles that of magnetically accreting white dwarfs \citep[polars and intermediate polars;][]{ezuka1999,muno2003,hong2009}. Although wind-accreting neutron stars in high-mass X-ray binaries (HMXB) may show similarly hard X-ray spectra, the low hydrogen absorption column density inferred from fitting the X-ray spectrum and the lack of any UVOT or catalogued optical/infrared counterpart within the $2.8''$ XRT positional uncertainty, renders a HMXB nature less likely. 

Based on its X-ray spectral properties and lack of a counterpart at longer wavelengths we tentatively classify \polar\ as a magnetically accreting white dwarf. The limited statistics of the XRT data does not allow us to investigate the timing properties of the source. Deeper X-ray observations may test our proposed source classification by searching for common features of magnetically accreting white dwarfs, such as prominent iron emission lines in the X-ray spectrum or coherent X-ray pulsations with a periodicity on the order of minutes to hours \citep[e.g,][]{ritter2003,muno2003,hong2009,kaur2010}. Follow-up observations at other wavelengths (optical/infrared) can further aid in the classification of this source.

\subsubsection{Main sequence stars}\label{subsubsec:ms_stars}

AX J1738.4--2902 is positionally coincident with an optical/infrared catalogued source. We also detect a likely counterpart at this position in our UVOT images. Based on the fact that we a find very soft X-ray spectrum with little absorption, we tentatively associate AX J1738.4--2902 with the catalogued optical/infrared object. As such, this \asca\ source is identified as a K0V pre-main sequence star (see Sect.~\ref{subsec:oir}). 

\begin{figure}
 \begin{center}
\includegraphics[width=8cm]{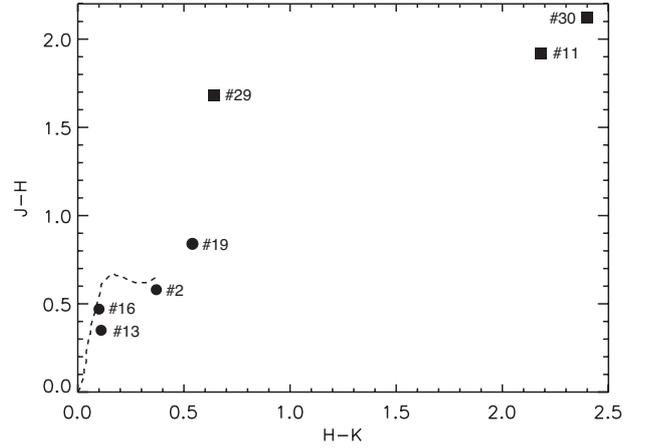}
    \end{center}
\caption[]{Approximate {\it 2MASS} colours for possible near-infrared counterparts of detected sources in our sample (magnitudes are not corrected for extinction). Circles represent sources that also have optical information, whereas squares indicate objects that are only detected in the infrared bands. The data points are labelled to indicate the corresponding source numbers (see Table~\ref{tab:oir}). The dashed line indicates the expectation for A- through M-type main-sequence stars. Correcting for extinction would move the data points down and towards the left in this diagram.
}
 \label{fig:ir}
\end{figure}

For another 7 of our XRT detections we find possible associations with X-ray/UV/optical/infrared/radio sources (see Tables~\ref{tab:oir} and~\ref{tab:class}). In Fig.~\ref{fig:ir} we plot the approximate {\it 2MASS} colours for the 6 possible infrared counterparts to detected \asca\ sources in our sample (magnitudes are not corrected for extinction).\footnote{Taking into account extinction would push the data points down and towards the left in the colour-colour diagram (Fig.~\ref{fig:ir}).} Circles represent sources which also have optical information, whereas squares denote those which are only detected in the infrared bands (see Table~\ref{tab:oir}). The dashed line in Fig.~\ref{fig:ir} indicates the expectation for A- through M-type main-sequence stars \citep[][]{bessell1988,tokunaga2000}. The labels to the data points correspond to the source numbers used in Table~\ref{tab:oir}.

We can see in Fig.~\ref{fig:ir} that three sources lie along the main-sequence line: AX J1504.6--5824, AX J1720.8--3710 and AX J1738.4--2902. Our proposed counterpart for the latter is indeed classified as a K0V star (see above). If AX J1504.6--5824 and AX J1720.8--3710 are related to the infrared objects located within their XRT error circles, both are likely main sequence stars as well. The low absorption column density ($N_H\lesssim3\times10^{21}~\nh$) inferred from fitting the X-ray spectrum of AX J1720.8--3710 makes this association likely. The statistics of the spectral data are limited, however, and the offset from the XRT position to the {\it 2MASS} source is relatively large (see Tables~\ref{tab:spec} and~\ref{tab:oir}). In the case of AX J1504.6--5824, there is also a considerable offset between the XRT position and the UV/optical/infrared object, whereas the X-ray spectrum appears relatively hard and highly absorbed (Tables~\ref{tab:spec} and~\ref{tab:oir}). This casts doubt on the possible classification as a foreground star.

We emphasise that the probability of a chance detection of an optical/infrared source within the XRT error circles is high (typically $\sim50\%$) as the source fields are crowded (see Sect.~\ref{subsec:oir}). It may not be a coincidence that the sources for which we find possible associations are often the ones that were detected at low significance in the XRT data and consequently have relatively large positional errors compared to the brighter XRT detections (see Table~\ref{tab:pos}). Our proposed possible associations may be confirmed or rejected by obtaining even more accurate X-ray positions (which is currently only possible with \chan), or by conducting dedicated follow-up observations at optical/infrared wavelengths.

\begin{figure}
 \begin{center}
\includegraphics[width=8cm]{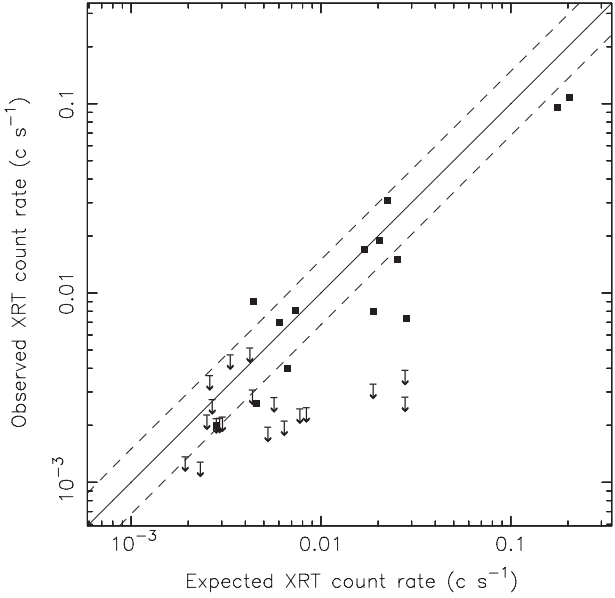}
    \end{center}
\caption[]{Observed versus expected XRT count rates for both detected and undetected sources in our sample (represented by squares and upper limit symbols, respectively). The solid line corresponds to a ratio of observed over expected count rates that equals unity, whereas the dashed lines represents $1\sigma$ deviations from this relation.
}
 \label{fig:rates}
\end{figure}

\begin{figure}
 \begin{center}
\includegraphics[width=8cm]{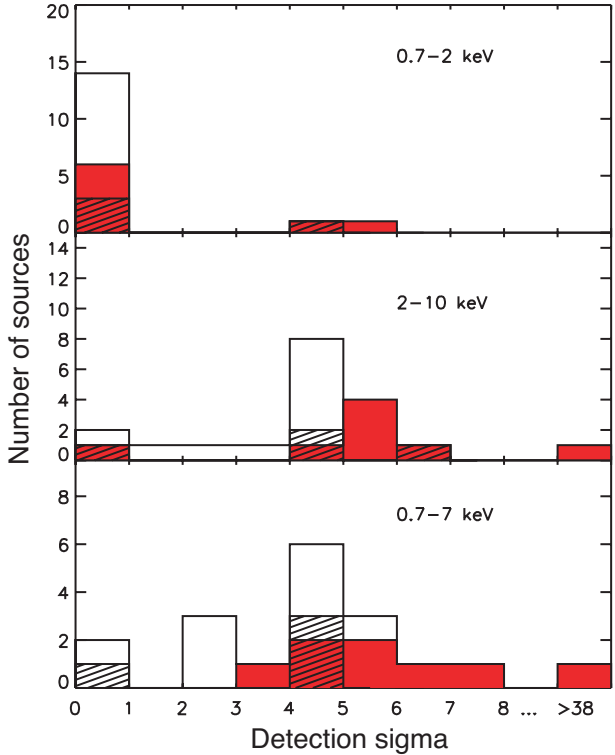}
    \end{center}
\caption[]{The source detection significances of the \asca\ Galactic plane survey are shown for the 3 energy bands investigated in that work \citep[][]{sugizaki01}. \swift/XRT non-detected sources are represented by the solid black line, the firmly detected sources are coloured red and tentative detections are indicated by hatched lines. These three different groups are overlaid in the graph (i.e., it is not cumulative).
}
 \label{fig:ascapar}
\end{figure}

\subsection{The nature of non-detected sources: a comparison with the \asca\ results}\label{subsec:nat_nondetect}

In Fig.~\ref{fig:rates}, we compare the observed and expected XRT count rates for our sample sources. This figure shows that to within 1$\sigma$ (dashed lines) only one source was brighter than predicted, while for 46\% of our sample the count rates lay below expectations. Among these fainter than expected sources are many of the XRT non-detections, implying either transient or highly variable sources, significant Eddington bias effects, extended rather than point-like sources or spurious detections in the \asca\ surveys. The Eddington bias, which boosts the true count rate of sources at or near the detection limit \citep[][]{eddington1913}, is likely to be at work in this sample as we are probing the faint end of the \asca\ catalogues. This idea is strengthened by the fact that we find the observed XRT count rates to be predominantly lower than expected. 

Without knowing the intrinsic source population we cannot estimate the magnitude of this effect, but we note that work on the \xmm\ hard band survey (2--10 keV) has shown that for the \xmm\ slew survey, the true count rate distribution peaks a factor of 2 below the observed count rate for detections with low numbers of counts \citep[][Warwick, R. et al., submitted to A\&A]{starling2011}. Three of our XRT non-detected sources have count rate upper limits that are a factor 5--10 lower than predicted from the \asca\ observations (see Table~\ref{tab:nondet}), which strongly suggests these are not faint, steady sources with count rates boosted by the Eddington bias, but are either variable (possibly transient) sources or spurious detections. Of these three, AX J1833.9--0822 was a firm 6$\sigma$ detection above 2 keV with \asca\ and AX J1836.3--0647 was a 5$\sigma$ detection in the 0.7--7 keV band dominated by its soft flux, while no similar information is available for AX J1751.1--2748. These detection significances disfavour a non-astrophysical, spurious nature. Various classes of X-ray sources (e.g., X-ray binaries, accreting white dwarfs, magnetars, stars, AGN) vary in intensity by a factor of a few. X-ray sources are generally denoted as transient when the variations are as large as $>2$ orders of magnitude. We thus tentatively identify AX J1751.1--2748, AX J1833.9--0822 and AX J1836.3--0647 as strongly variable or transient X-ray sources, although we cannot classify them further.

While it is not possible to classify the sources \swift\ has not detected, we may learn something of their nature from further examining the details of their \asca\ measurements. A total of 26 objects in our sample were drawn from the \asca\ Galactic plane survey, which were evaluated in 3 energy bands (0.7--2 keV, 2--10 keV and 0.7--7 keV) and assigned a detection significance \citep[][]{sugizaki01}. We plot these values in Fig.~\ref{fig:ascapar}, distinguishing between the XRT non-detections (solid black lines), detections (coloured red) and tentative detections (hatched lines). It can be seen that just 3 sources were significantly detected in the soft energy band (0.7--2 keV), and these were also detected both in other \asca\ bands and later with the XRT. This is unsurprising given the high absorbing columns in these directions (see Tables~\ref{tab:spec} and \ref{tab:nondet}). This causes the broad band (0.7--7 keV) to largely follow the results of the hard band (2--10 keV), where XRT-detected sources generally represent the higher-significance \asca\ detections while XRT non-detections are found below 5--6$\sigma$. Our tentative XRT detections span a wider range of \asca\ detection significances, and are concentrated in the 0.7--7 keV band over the 4--5$\sigma$ range.

It is of note that Galactic plane survey sources with $>$4$\sigma$ detections in all 3 \asca\ energy bands number just two, AX J1538.3--5541 and AX J1651.0--4403, both of which are well detected with \swift/XRT (see Sect.~\ref{subsubsec:porb} for further details of AX J1538.3--5541). There are 3 hard-band only sources: 1 was tentatively detected with the XRT and the other 2 were not. Broad-band only detections generally have the fewest total counts and are therefore most likely to be spurious. We have two, AX J1835.1--0806 and AX J1857.3+0247. Both went undetected with the XRT adding further uncertainty to assignment of an astrophysical nature to these sources. We thus tentatively mark the two as spurious \asca\ detections. However, we note that their XRT count rate limits are not more than a factor of 2 below the expected rates. This is certainly possible within the Eddington bias effect and a large number of sources are variable at this level. Furthermore, our expected XRT count rates are overestimated if the sources are extended rather than point-like. As mentioned in Sect.~\ref{subsec:hard}, AX J1857.3+0247 may possibly be related to an extended X-ray source.

Spectral analysis was attempted for those sources with \asca\ detections above 5$\sigma$ in both \citet{sugizaki01} and \citet{sakano02}, over the full energy range of 0.7--10 keV and adopting an absorbed power law model, although only \citet{sugizaki01} provide error estimates on the parameters. For the five firm XRT detections with well-measured \asca\ spectral parameters, $N_H$ and $\Gamma$ are consistent within the 90\% uncertainties with one exception: AX J1538.3--5541. For this source $N_H$ is consistent between the \asca\ and \swift/XRT averaged spectra, while the photon index $\Gamma$ is just consistent at the 3$\sigma$ level possibly having been harder during the \asca\ observation. We found this source to be highly time variable from multiple XRT observations, and possible spectral evolution is investigated in Sect.~\ref{subsubsec:porb}. The overall consistency between spectral measurements validates our use of the \asca\ spectra to make count rate predictions for XRT as used in Tables~\ref{tab:spec} and~\ref{tab:nondet}.

In conclusion, the 31\% of sample sources firmly detected with {\it Swift}/XRT were generally found with greater confidence in the \asca\ Galactic plane and Galactic centre surveys than the XRT non-detected sources. Very few sources were detected below 2 keV with \asca, which is likely due to high foreground absorbing columns (see Tables~\ref{tab:spec} and~\ref{tab:nondet}). However, the greatest deviation from the expected count rate among the XRT non-detections corresponds to three sources that were largely well-detected with \asca, strengthening the probability that they are astrophysical sources of variable or transient nature. We identify two sources that most likely have been spurious \asca-detections given that these were weakly detected only in the full \asca\ energy band and not detected with XRT. For sources well-detected with both \asca\ and XRT we find their spectra to be unchanged within the given errors between the two epochs.

\subsection{Summary}\label{subsec:summary}
Our sample of 35 unclassified \asca\ sources includes one confirmed and one candidate X-ray binary, and likely one magnetically accreting white dwarf. We assign these classifications based on their X-ray spectral and temporal properties. Furthermore, we used the improved XRT positions of the 16 sources detected in our sample to search for possible associations with catalogued sources at X-ray, UV, optical, infrared and radio wavelengths. This results in the identifications of possible counterparts for 8 of our targets, amongst which are 3 likely main sequence stars (see Table~\ref{tab:class}). The probability of an optical/infrared chance detection is, however, considerable in these crowded fields ($\sim50\%$). 

By assessing any possible long-term variability and the reported \asca\ properties of the 19 sources that were not detected by XRT, we identify two potentially spurious \asca-detections amongst them and three X-ray sources that appear to be variable or transient (see Table~\ref{tab:class}). A substantial fraction of the XRT non-detections are expected to be due to the Eddington bias, and might thus involve weak, persistent X-ray sources with intensities that lie near the detection limit of our \swift/XRT observations. 

With our improved $\sim2-4''$ X-ray positions obtained for the 16 XRT-detected sources dedicated follow-up observations at different wavelengths become feasible. These have the potential to further classify these faint X-ray sources. As can be seen in Table~\ref{tab:class}, our study shows that the unclassified \asca\ sources harbour a variety of astrophysical objects.

\begin{table*}
\begin{threeparttable}[t]
\begin{center}
\caption[]{{Proposed source associations/classifications.}}
\begin{tabular}{l l l}
\toprule
$\#$ & Name & Comments   \\
\midrule
\multicolumn{3}{l}{{\bf Firm \swift/XRT detections}} \\ 
2 & AX J1504.6--5824 & Possible association with an optical/infrared object: main sequence star \\
5 & AX J1538.3--5541 & Candidate X-ray binary, most likely an LMXB \\
7 & \polar & Candidate accreting magnetised white dwarf\\
16 & AX J1738.4--2902 & Likely associated with 1RXS J173826.7--290140/2MASS J17382620--2901494: pre-main sequence star (K0V) \\
19 & AX J1742.6--2901 & Possible association with 2RXP J174241.8--290215, optical/infrared/radio object \\
22 & AX J1754.2--2754 & Confirmed neutron star LMXB, proposed ultra-compact X-ray binary \\
29 & AX J1846.0--0231 & Possible association with an infrared object\\
\midrule
\multicolumn{3}{l}{{\bf Tentative \swift/XRT detections}} \\ 
11 & AX J1717.2--3718 & Possible association with an infrared object \\
13 & AX J1720.8--3710 & Possibly related to 1RXS 172051.9--371033, UV/optical/infrared object: main sequence star \\
17 & AX J1739.5--2730 & Possible association with an optical object \\
30 & AX J1846.1--0239 & Possible association with an optical/infrared object \\
\midrule
\multicolumn{3}{l}{{\bf \swift/XRT non-detections}} \\ 
20 & AX J1751.1--2748 & Strongly variable/transient X-ray source \\
25 & AX J1833.9--0822 & Strongly variable/transient X-ray source \\
27 & AX J1835.1--0806 & Potential spurious \asca\ detection \\
28 & AX J1836.3--0647 & Strongly variable/transient X-ray source \\
34 & AX J1857.3+0247 & Potential spurious \asca\ detection \\
\bottomrule
\end{tabular}
\label{tab:class}
\begin{tablenotes}
\item[]
\end{tablenotes}
\end{center}
\end{threeparttable}
\end{table*}

\begin{acknowledgements}
The authors acknowledge the use of public data from the \textit{Swift} data archive and data supplied by the UK \swift\ Science Data Center at the University of Leicester. Swift is supported at PSU by NASA contract NAS5-00136. This research has made use of the XRT Data Analysis Software (XRTDAS) developed under the responsibility of the ASI Science Data Center (ASDC), Italy. This work was supported by the Netherlands Organisation for Scientific Research (NWO) and the Netherlands Research School for Astronomy (NOVA). ND is supported by NASA through Hubble Postdoctoral Fellowship grant number HST-HF-51287.01-A from the Space Telescope Science Institute, which is operated by the Association of Universities for Research in Astronomy, Incorporated, under NASA contract NAS5-26555. RLCS is supported by a Royal Society Fellowship. PAE, APB and KLP acknowledge support by the UK Space Agency (UKSA). RW is supported by a European Research Council (ERC) starting grant. ND acknowledges the hospitality of the University of Leicester where most of this work was carried out.
\end{acknowledgements}

\bibliographystyle{aa}
\bibliography{thesis}

\end{document}